\RequirePackage{fix-cm}
\documentclass[twocolumn,epjc3]{svjour3} 
\RequirePackage[T1]{fontenc}

\smartqed

\RequirePackage{graphicx}
\RequirePackage{mathptmx}     
\RequirePackage{bm}
\RequirePackage{flushend}
\RequirePackage[numbers,sort&compress]{natbib}

\usepackage{paralist}
\usepackage{siunitx}
\usepackage{xcolor}
\usepackage[version=4]{mhchem}
\usepackage{mathtools}

\RequirePackage[colorlinks,citecolor=blue,urlcolor=blue,linkcolor=blue]{hyperref}

\journalname{Eur. Phys. J. C}

\usepackage[normalem]{ulem}

\begin{document}

\onecolumn

\title{Simulation of the background from $^{13}$C$(\alpha,\,n)^{16}$O reaction \\ in the JUNO scintillator}

\subtitle{JUNO Collaboration\thanksref{email_jpc}}

\author{
Thomas Adam\thanksref{37} \and 
Kai Adamowicz\thanksref{40} \and
Shakeel Ahmad\thanksref{57} \and
Rizwan Ahmed\thanksref{57} \and
Sebastiano Aiello\thanksref{47} \and
Fengpeng An\thanksref{16} \and
Costas Andreopoulos\thanksref{65} \and
Giuseppe Andronico\thanksref{47} \and
Nikolay Anfimov\thanksref{58} \and
Vito Antonelli\thanksref{49} \and
Tatiana Antoshkina\thanksref{58} \and
Jo\~{a}o Pedro Athayde Marcondes de Andr\'{e}\thanksref{37} \and
Didier Auguste\thanksref{35} \and
Weidong Bai\thanksref{16} \and
Nikita Balashov\thanksref{58} \and 
Andrea Barresi\thanksref{50} \and
Davide Basilico\thanksref{49} \and
Eric Baussan\thanksref{37} \and
Marco Beretta\thanksref{49} \and
Antonio Bergnoli\thanksref{52} \and
Nikita Bessonov\thanksref{58}  \and
Daniel Bick\thanksref{41} \and
Lukas Bieger\thanksref{46} \and
Svetlana Biktemerova\thanksref{58} \and
Thilo Birkenfeld\thanksref{40} \and
Simon Blyth\thanksref{8} \and
Anastasia Bolshakova\thanksref{58} \and
Mathieu Bongrand\thanksref{39} \and
Matteo Borghesi\thanksref{50} \and
Dominique Breton\thanksref{35} \and
Augusto Brigatti\thanksref{49} \and
Riccardo Brugnera\thanksref{53} \and
Riccardo Bruno\thanksref{47} \and
Marcel B\"{u}chner\thanksref{43} \and
Antonio Budano\thanksref{56} \and
Jose Busto\thanksref{38} \and
Anatael Cabrera\thanksref{35} \and
Barbara Caccianiga\thanksref{49} \and
Hao Cai\thanksref{26} \and
Xiao Cai\thanksref{8} \and
Yanke Cai\thanksref{8} \and
Zhiyan Cai\thanksref{8} \and
St\'{e}phane Callier\thanksref{36} \and
Steven Calvez\thanksref{39} \and
Antonio Cammi\thanksref{51,present} \and
Chuanya Cao\thanksref{8} \and
Guofu Cao\thanksref{8} \and
Jun Cao\thanksref{8} \and
Yaoqi Cao\thanksref{66,65} \and
Rossella Caruso\thanksref{47} \and
C\'{e}dric Cerna\thanksref{36} \and
Vanessa Cerrone\thanksref{53} \and
Jinfan Chang\thanksref{8} \and
Yun Chang\thanksref{31} \and
Auttakit Chatrabhuti\thanksref{62} \and
Chao Chen\thanksref{8} \and
Guoming Chen\thanksref{22} \and
Jiahui Chen\thanksref{8} \and
Jian Chen\thanksref{16} \and
Jing Chen\thanksref{16} \and
Junyou Chen\thanksref{22} \and
Pingping Chen\thanksref{14} \and
Shaomin Chen\thanksref{10} \and
Shiqiang Chen\thanksref{21} \and
Xin Chen\thanksref{21,8} \and
Yiming Chen\thanksref{8} \and
Yixue Chen\thanksref{9} \and
Yu Chen\thanksref{16} \and
Ze Chen\thanksref{45,43} \and
Zhangming Chen\thanksref{23} \and
Zhiyuan Chen\thanksref{8} \and
Jie Cheng\thanksref{9} \and
Yaping Cheng\thanksref{7} \and
Yu Chin Cheng\thanksref{32} \and
Alexander Chepurnov\thanksref{60,59} \and
Alexey Chetverikov\thanksref{58} \and
Davide Chiesa\thanksref{50} \and
Pietro Chimenti\thanksref{3} \and
Po-Lin Chou\thanksref{30} \and
Ziliang Chu\thanksref{8} \and
Artem Chukanov\thanksref{58} \and
G\'{e}rard Claverie\thanksref{36} \and
Catia Clementi\thanksref{54} \and
Barbara Clerbaux\thanksref{2} \and
Claudio Coletta\thanksref{50} \and
Selma Conforti Di Lorenzo\thanksref{36} \and
Simon Csakli\thanksref{44} \and
Chenyang Cui\thanksref{8} \and
Olivia Dalager\thanksref{67} \and
Christophe De La Taille\thanksref{36} \and
Zhi Deng\thanksref{10} \and
Ziyan Deng\thanksref{8} \and
Xiaoyu Ding\thanksref{20} \and
Xuefeng Ding\thanksref{8} \and
Yayun Ding\thanksref{8} \and
Bayu Dirgantara\thanksref{64} \and
Carsten Dittrich\thanksref{44} \and
Sergey Dmitrievsky\thanksref{58} \and
David Doerflinger\thanksref{44} \and
Dmitry Dolzhikov\thanksref{58} \and
Haojie Dong\thanksref{8} \and
Jianmeng Dong\thanksref{10} \and
Evgeny Doroshkevich\thanksref{59} \and
Marcos Dracos\thanksref{37} \and
Fr\'{e}d\'{e}ric Druillole\thanksref{36} \and
Ran Du\thanksref{8} \and
Shuxian Du\thanksref{29} \and
Yujie Duan\thanksref{26} \and
Katherine Dugas\thanksref{67} \and
Stefano Dusini\thanksref{52} \and
Hongyue Duyang\thanksref{20} \and
Jessica Eck\thanksref{46} \and
Timo Enqvist\thanksref{34} \and
Andrea Fabbri\thanksref{56} \and
Ulrike Fahrendholz\thanksref{44} \and
Lei Fan\thanksref{8} \and
Jian Fang\thanksref{8} \and
Wenxing Fang\thanksref{8} \and
Dmitry Fedoseev\thanksref{58} \and
Li-Cheng Feng\thanksref{30} \and
Qichun Feng\thanksref{17} \and
Federico Ferraro\thanksref{49} \and
Daniela Fetzer\thanksref{43} \and
Marcellin Fotz\'{e}\thanksref{37} \and
Am\'{e}lie Fournier\thanksref{36} \and
Aaron Freegard\thanksref{23} \and
Feng Gao\thanksref{2} \and
Alberto Garfagnini\thanksref{53} \and
Arsenii Gavrikov\thanksref{53} \and
Marco Giammarchi\thanksref{49} \and
Nunzio Giudice\thanksref{47} \and
Maxim Gonchar\thanksref{58} \and
Guanghua Gong\thanksref{10} \and
Hui Gong\thanksref{10} \and
Yuri Gornushkin\thanksref{58} \and
Marco Grassi\thanksref{53} \and
Maxim Gromov\thanksref{60,58} \and
Vasily Gromov\thanksref{58} \and
Minghao Gu\thanksref{8} \and
Xiaofei Gu\thanksref{29} \and
Yu Gu\thanksref{15} \and
Mengyun Guan\thanksref{8} \and
Yuduo Guan\thanksref{8} \and
Nunzio Guardone\thanksref{47} \and
Rosa Maria Guizzetti\thanksref{53} \and
Cong Guo\thanksref{8} \and
Wanlei Guo\thanksref{8} \and
Caren Hagner\thanksref{41} \and
Hechong Han\thanksref{8} \and
Ran Han\thanksref{7} \and
Yang Han\thanksref{16} \and
Jinhong He\thanksref{26} \and
Miao He\thanksref{8} \and
Wei He\thanksref{8} \and
Xinhai He\thanksref{8} \and
Ziou He\thanksref{66,65} \and
Tobias Heinz\thanksref{46} \and
Patrick Hellmuth\thanksref{36} \and
Yuekun Heng\thanksref{8} \and
YuenKeung Hor\thanksref{16} \and
Shaojing Hou\thanksref{8} \and
Yee Hsiung\thanksref{32} \and
Bei-Zhen Hu\thanksref{32} \and
Hang Hu\thanksref{16} \and
Jun Hu\thanksref{8} \and
Tao Hu\thanksref{8} \and
Yuxiang Hu\thanksref{8} \and
Guihong Huang\thanksref{19} \and
Jinhao Huang\thanksref{8} \and
Junting Huang\thanksref{23} \and
Kaixuan Huang\thanksref{16} \and
Shengheng Huang\thanksref{19} \and
Tao Huang\thanksref{16} \and
Xin Huang\thanksref{8} \and
Xingtao Huang\thanksref{20} \and
Yongbo Huang\thanksref{22} \and
Jiaqi Hui\thanksref{23} \and
Lei Huo\thanksref{17} \and
C\'{e}dric Huss\thanksref{36} \and
Safeer Hussain\thanksref{57} \and
Leonard Imbert\thanksref{39} \and
Ara Ioannisian\thanksref{1} \and
Adrienne Jacobi\thanksref{67} \and
Arshak Jafar\thanksref{43} \and
Beatrice Jelmini\thanksref{53} \and
Xiangpan Ji\thanksref{25} \and
Xiaolu Ji\thanksref{8} \and
Huihui Jia\thanksref{25} \and
Junji Jia\thanksref{26} \and
Cailian Jiang\thanksref{21} \and
Wei Jiang\thanksref{8} \and
Xiaoshan Jiang\thanksref{8} \and
Xiaozhao Jiang\thanksref{8} \and
Yijian Jiang\thanksref{9} \and
Yixuan Jiang\thanksref{8} \and
Xiaoping Jing\thanksref{8} \and
C\'{e}cile Jollet\thanksref{36} \and
Li Kang\thanksref{14} \and
Rebin Karaparabil\thanksref{37} \and
Narine Kazarian\thanksref{1} \and
Ali Khan\thanksref{57} \and
Amina Khatun\thanksref{2,61} \and
Khanchai Khosonthongkee\thanksref{64} \and
Denis Korablev\thanksref{58} \and
Konstantin Kouzakov\thanksref{60} \and
Alexey Krasnoperov\thanksref{58} \and
Sergey Kuleshov\thanksref{5} \and
Sindhujha Kumaran\thanksref{67} \and
Nikolay Kutovskiy\thanksref{58} \and
Loïc Labit\thanksref{36} \and
Tobias Lachenmaier\thanksref{46} \and
Haojing Lai\thanksref{23} \and
Cecilia Landini\thanksref{49} \and
S\'{e}bastien Leblanc\thanksref{36} \and
Matthieu Lecocq\thanksref{36} \and
Frederic Lefevre\thanksref{39} \and
Ruiting Lei\thanksref{14} \and
Rupert Leitner\thanksref{33} \and
Jason Leung\thanksref{30} \and
Demin Li\thanksref{29} \and
Fei Li\thanksref{8} \and
Fule Li\thanksref{10} \and
Gaosong Li\thanksref{8} \and
Hongjian Li\thanksref{8} \and
Huang Li\thanksref{8} \and
Jiajun Li\thanksref{16} \and
Min Li\thanksref{37} \and
Nan Li\thanksref{12} \and
Qingjiang Li\thanksref{12} \and
Ruhui Li\thanksref{8} \and
Rui Li\thanksref{23} \and
Shanfeng Li\thanksref{14} \and
Tao Li\thanksref{16} \and
Teng Li\thanksref{20} \and
Weidong Li\thanksref{8,11} \and
Xiaonan Li\thanksref{8} \and
Yi Li\thanksref{14} \and
Yichen Li\thanksref{8} \and
Yifan Li\thanksref{8} \and
Yufeng Li\thanksref{8} \and
Zhaohan Li\thanksref{8} \and
Zhibing Li\thanksref{16} \and
Zi-Ming Li\thanksref{29} \and
Zonghai Li\thanksref{26} \and
An-An Liang\thanksref{30} \and
Jiajun Liao\thanksref{16} \and
Minghua Liao\thanksref{16} \and
Yilin Liao\thanksref{23} \and
Ayut Limphirat\thanksref{64} \and
Bo-Chun Lin\thanksref{30} \and
Guey-Lin Lin\thanksref{30} \and
Shengxin Lin\thanksref{14} \and
Tao Lin\thanksref{8} \and
Xianhao Lin\thanksref{21} \and
Xingyi Lin\thanksref{22} \and
Jiajie Ling\thanksref{16} \and
Xin Ling\thanksref{18} \and
Ivano Lippi\thanksref{52} \and
Caimei Liu\thanksref{8} \and
Fang Liu\thanksref{9} \and
Fengcheng Liu\thanksref{9} \and
Haidong Liu\thanksref{29} \and
Haotian Liu\thanksref{26} \and
Hongbang Liu\thanksref{22} \and
Hongjuan Liu\thanksref{18} \and
Hongtao Liu\thanksref{16} \and
Hongyang Liu\thanksref{8} \and
Jianglai Liu\thanksref{23,24} \and
Jiaxi Liu\thanksref{8} \and
Jinchang Liu\thanksref{8} \and
Kainan Liu\thanksref{19} \and
Min Liu\thanksref{18} \and
Qian Liu\thanksref{11} \and
Runxuan Liu\thanksref{45,40} \and
Shenghui Liu\thanksref{8} \and
Shulin Liu\thanksref{8} \and
Xiaowei Liu\thanksref{16} \and
Xiwen Liu\thanksref{22} \and
Xuewei Liu\thanksref{10} \and
Yankai Liu\thanksref{27} \and
Zhen Liu\thanksref{8} \and
Lorenzo Loi\thanksref{51} \and
Alexey Lokhov\thanksref{60,59} \and
Paolo Lombardi\thanksref{49} \and
Claudio Lombardo\thanksref{47} \and
Kai Loo\thanksref{34} \and
Haoqi Lu\thanksref{8} \and
Junguang Lu\thanksref{8} \and
Meishu Lu\thanksref{44,43} \and
Peizhi Lu\thanksref{16} \and
Shuxiang Lu\thanksref{29} \and
Xianguo Lu\thanksref{66} \and
Bayarto Lubsandorzhiev\thanksref{59} \and
Sultim Lubsandorzhiev\thanksref{59} \and
Livia Ludhova\thanksref{45,43} \and
Arslan Lukanov\thanksref{59} \and
Fengjiao Luo\thanksref{18} \and
Guang Luo\thanksref{16} \and
Jianyi Luo\thanksref{16} \and
Shu Luo\thanksref{28} \and
Wuming Luo\thanksref{8} \and
Xiaojie Luo\thanksref{8} \and
Vladimir Lyashuk\thanksref{59} \and
Bangzheng Ma\thanksref{20} \and
Bing Ma\thanksref{29} \and
Qiumei Ma\thanksref{8} \and
Si Ma\thanksref{8} \and
Wing Yan Ma\thanksref{20} \and
Xiaoyan Ma\thanksref{8} \and
Xubo Ma\thanksref{9} \and
Jihane Maalmi\thanksref{35} \and
Jingyu Mai\thanksref{16} \and
Marco Malabarba\thanksref{45,43} \and
Yury Malyshkin\thanksref{45,43} \and
Roberto Carlos Mandujano\thanksref{67} \and
Fabio Mantovani\thanksref{48} \and
Xin Mao\thanksref{7} \and
Stefano M. Mari\thanksref{56} \and
Agnese Martini\thanksref{55} \and
Matthias Mayer\thanksref{44} \and
Davit Mayilyan\thanksref{1} \and
Yue Meng\thanksref{23} \and
Anselmo Meregaglia\thanksref{36} \and
Lino Miramonti\thanksref{49} \and
Marta Colomer Molla\thanksref{2} \and
Michele Montuschi\thanksref{48} \and
Cristobal Morales Reveco\thanksref{45,40,43} \and
Iwan Morton-Blake\thanksref{24} \and
Massimiliano Nastasi\thanksref{50} \and
Dmitry V. Naumov\thanksref{58} \and
Elena Naumova\thanksref{58} \and
Igor Nemchenok\thanksref{58} \and
Elisabeth Neuerburg\thanksref{40} \and
Minh Thuan Nguyen Thi\thanksref{30} \and
Alexey Nikolaev\thanksref{60} \and
Feipeng Ning\thanksref{8} \and
Zhe Ning\thanksref{8} \and
Yujie Niu\thanksref{8} \and
Hiroshi Nunokawa\thanksref{4} \and
Lothar Oberauer\thanksref{44,43} \and
Juan Pedro Ochoa-Ricoux\thanksref{67,5} \and
Sebastian Olivares\thanksref{6} \and
Alexander Olshevskiy\thanksref{58} \and
Domizia Orestano\thanksref{56} \and
Fausto Ortica\thanksref{54} \and
Rainer Othegraven\thanksref{43} \and
Yifei Pan\thanksref{16} \and
Alessandro Paoloni\thanksref{55} \and
George Parker\thanksref{43} \and
Yatian Pei\thanksref{8} \and
Luca Pelicci\thanksref{49} \and
Anguo Peng\thanksref{18} \and
Yu Peng\thanksref{8} \and
Zhaoyuan Peng\thanksref{8} \and
Elisa Percalli\thanksref{49} \and
Willy Perrin\thanksref{37} \and
Fr\'{e,}d\'{e,}ric Perrot\thanksref{36} \and
Pierre-Alexandre Petitjean\thanksref{2} \and
Fabrizio Petrucci\thanksref{56} \and
Oliver Pilarczyk\thanksref{43} \and
Artyom Popov\thanksref{60} \and
Pascal Poussot\thanksref{37} \and
Ezio Previtali\thanksref{50} \and
Fazhi Qi\thanksref{8} \and
Ming Qi\thanksref{21} \and
Xiaohui Qi\thanksref{8} \and
Sen Qian\thanksref{8} \and
Xiaohui Qian\thanksref{8} \and
Zhonghua Qin\thanksref{8} \and
Shoukang Qiu\thanksref{18} \and
Manhao Qu\thanksref{29} \and
Zhenning Qu\thanksref{8} \and
Gioacchino Ranucci\thanksref{49} \and
Thomas Raymond\thanksref{37} \and
Alessandra Re\thanksref{49} \and
Abdel Rebii\thanksref{36} \and
Mariia Redchuk\thanksref{52} \and
Bin Ren\thanksref{14} \and
Yuhan Ren\thanksref{8} \and
Barbara Ricci\thanksref{48} \and
Komkrit Rientong\thanksref{62} \and
Mariam Rifai\thanksref{45,40,43} \and
Mathieu Roche\thanksref{36} \and
Narongkiat Rodphai\thanksref{8} \and
Fernanda de Faria Rodrigues\thanksref{8} \and
Aldo Romani\thanksref{54} \and
Bed\v{r}ich Roskovec\thanksref{33} \and
Arseniy Rybnikov\thanksref{58} \and
Andrey Sadovsky\thanksref{58} \and
Paolo Saggese\thanksref{49} \and
Deshan Sandanayake\thanksref{37} \and
Anut Sangka\thanksref{63} \and
Giuseppe Sava\thanksref{47} \and
Utane Sawangwit\thanksref{63} \and
Michaela Schever\thanksref{40} \and
C\'{e}dric Schwab\thanksref{37} \and
Konstantin Schweizer\thanksref{44} \and
Alexandr Selyunin\thanksref{58} \and
Andrea Serafini\thanksref{53} \and
Mariangela Settimo\thanksref{39} \and
Junyu Shao\thanksref{8} \and
Vladislav Sharov\thanksref{58} \and
Hangyu Shi\thanksref{16} \and
Hexi Shi\thanksref{45,42} \and
Jingyan Shi\thanksref{8} \and
Yanan Shi\thanksref{8} \and
Vitaly Shutov\thanksref{58} \and
Andrey Sidorenkov\thanksref{59} \and
Fedor \v{S}imkovic\thanksref{61} \and
Apeksha Singhal\thanksref{45,43} \and
Chiara Sirignano\thanksref{53} \and
Jaruchit Siripak\thanksref{64} \and
Monica Sisti\thanksref{50} \and
Oleg Smirnov\thanksref{58} \and
Sergey Sokolov\thanksref{58} \and
Julanan Songwadhana\thanksref{64} \and
Boonrucksar Soonthornthum\thanksref{63} \and
Albert Sotnikov\thanksref{58} \and
Warintorn Sreethawong\thanksref{64} \and
Achim Stahl\thanksref{40} \and
Luca Stanco\thanksref{52} \and
Elia Stanescu Farilla\thanksref{56} \and
Konstantin Stankevich\thanksref{60} \and
Hans Steiger\thanksref{44,43} \and
Jochen Steinmann\thanksref{40} \and
Tobias Sterr\thanksref{46} \and
Raphael Stock\thanksref{44,43} \and
Virginia Strati\thanksref{48} \and
Mikhail Strizh\thanksref{60,59} \and
Alexander Studenikin\thanksref{60} \and
Aoqi Su\thanksref{29} \and
Jun Su\thanksref{16} \and
Guangbao Sun\thanksref{26} \and
Mingxia Sun\thanksref{8} \and
Shifeng Sun\thanksref{9} \and
Xilei Sun\thanksref{8} \and
Yongzhao Sun\thanksref{8} \and
Zhengyang Sun\thanksref{24} \and
Narumon Suwonjandee\thanksref{62} \and
Akira Takenaka\thanksref{24} \and
Xiaohan Tan\thanksref{20} \and
Jian Tang\thanksref{16} \and
Jingzhe Tang\thanksref{22} \and
Qiang Tang\thanksref{16} \and
Quan Tang\thanksref{18} \and
Xiao Tang\thanksref{8} \and
Vidhya Thara Hariharan\thanksref{41} \and
Yuxin Tian\thanksref{24} \and
Igor Tkachev\thanksref{59} \and
Tomas Tmej\thanksref{33} \and
Marco Danilo Claudio Torri\thanksref{49} \and
Andrea Triossi\thanksref{53} \and
Wladyslaw Trzaska\thanksref{34} \and
Yu-Chen Tung\thanksref{32} \and
Cristina Tuve\thanksref{47} \and
Nikita Ushakov\thanksref{59} \and
Carlo Venettacci\thanksref{56} \and
Giuseppe Verde\thanksref{47} \and
Maxim Vialkov\thanksref{60} \and
Benoit Viaud\thanksref{39} \and
Cornelius Moritz Vollbrecht\thanksref{45,40} \and
Vit Vorobel\thanksref{33} \and
Dmitriy Voronin\thanksref{59} \and
Lucia Votano\thanksref{55} \and
Caishen Wang\thanksref{14} \and
Chung-Hsiang Wang\thanksref{31} \and
En Wang\thanksref{29} \and
Hanwen Wang\thanksref{8} \and
Jiabin Wang\thanksref{20} \and
Jun Wang\thanksref{16} \and
Li Wang\thanksref{29,8} \and
Meng Wang\thanksref{18} \and
Meng Wang\thanksref{20} \and
Mingyuan Wang\thanksref{8} \and
Qianchuan Wang\thanksref{26} \and
Ruiguang Wang\thanksref{8} \and
Sibo Wang\thanksref{8} \and
Tianhong Wang\thanksref{17} \and
Wei Wang\thanksref{16} \and
Wenshuai Wang\thanksref{8} \and
Xi Wang\thanksref{12} \and
Yangfu Wang\thanksref{8} \and
Yaoguang Wang\thanksref{20} \and
Yi Wang\thanksref{8} \and
Yi Wang\thanksref{10} \and
Yifang Wang\thanksref{8} \and
Yuanqing Wang\thanksref{10} \and
Yuyi Wang\thanksref{10} \and
Zhe Wang\thanksref{10} \and
Zheng Wang\thanksref{8} \and
Zhimin Wang\thanksref{8} \and
Apimook Watcharangkool\thanksref{63} \and
Wei Wei\thanksref{8} \and
Wei Wei\thanksref{20} \and
Yadong Wei\thanksref{14} \and
Yuehuan Wei\thanksref{16} \and
Zhengbao Wei\thanksref{22} \and
Liangjian Wen\thanksref{8} \and
Jun Weng\thanksref{10} \and
Christopher Wiebusch\thanksref{40} \and
Rosmarie Wirth\thanksref{41} \and
Bi Wu\thanksref{16} \and
Chengxin Wu\thanksref{16} \and
Diru Wu\thanksref{8} \and
Qun Wu\thanksref{20} \and
Yinhui Wu\thanksref{8} \and
Yiyang Wu\thanksref{10} \and
Zhaoxiang Wu\thanksref{8} \and
Zhi Wu\thanksref{8} \and
Michael Wurm\thanksref{43} \and
Jacques Wurtz\thanksref{37} \and
Dongmei Xia\thanksref{13} \and
Shishen Xian\thanksref{24} \and
Ziqian Xiang\thanksref{23} \and
Fei Xiao\thanksref{8} \and
Pengfei Xiao\thanksref{8} \and
Xiang Xiao\thanksref{16} \and
Wei-Jun Xie\thanksref{30} \and
Xiaochuan Xie\thanksref{22} \and
Yijun Xie\thanksref{8} \and
Yuguang Xie\thanksref{8} \and
Zhao Xin\thanksref{8} \and
Zhizhong Xing\thanksref{8} \and
Benda Xu\thanksref{10} \and
Cheng Xu\thanksref{18} \and
Donglian Xu\thanksref{24,23} \and
Fanrong Xu\thanksref{15} \and
Jiayang Xu\thanksref{8} \and
Jilei Xu\thanksref{8} \and
Jinghuan Xu\thanksref{22} \and
Meihang Xu\thanksref{8} \and
Shiwen Xu\thanksref{8} \and
Xunjie Xu\thanksref{8} \and
Yin Xu\thanksref{25} \and
Yu Xu\thanksref{16} \and
Jingqin Xue\thanksref{8} \and
Baojun Yan\thanksref{8} \and
Qiyu Yan\thanksref{11,66} \and
Taylor Yan\thanksref{64} \and
Xiongbo Yan\thanksref{8} \and
Yupeng Yan\thanksref{64} \and
Changgen Yang\thanksref{8} \and
Chengfeng Yang\thanksref{16} \and
Fengfan Yang\thanksref{8} \and
Jie Yang\thanksref{29} \and
Lei Yang\thanksref{14} \and
Pengfei Yang\thanksref{16} \and
Xiaoyu Yang\thanksref{8} \and
Yifan Yang\thanksref{2} \and
Yixiang Yang\thanksref{8} \and
Zekun Yang\thanksref{20} \and
Haifeng Yao\thanksref{8} \and
Jiaxuan Ye\thanksref{8} \and
Mei Ye\thanksref{8} \and
Ziping Ye\thanksref{24} \and
Fr\'{e}d\'{e}ric Yermia\thanksref{39} \and
Jilong Yin\thanksref{8} \and
Weiqing Yin\thanksref{8} \and
Xiaohao Yin\thanksref{16} \and
Zhengyun You\thanksref{16} \and
Boxiang Yu\thanksref{8} \and
Chiye Yu\thanksref{14} \and
Chunxu Yu\thanksref{25} \and
Hongzhao Yu\thanksref{8} \and
Peidong Yu\thanksref{8} \and
Xianghui Yu\thanksref{25} \and
Zeyuan Yu\thanksref{8} \and
Zezhong Yu\thanksref{8} \and
Cenxi Yuan\thanksref{16} \and
Chengzhuo Yuan\thanksref{8} \and
Zhaoyang Yuan\thanksref{8} \and
Zhenxiong Yuan\thanksref{10} \and
Noman Zafar\thanksref{57} \and
Kirill Zamogilnyi\thanksref{60} \and
Jilberto Zamora\thanksref{6} \and
Vitalii Zavadskyi\thanksref{58} \and
Fanrui Zeng\thanksref{20} \and
Shan Zeng\thanksref{8} \and
Tingxuan Zeng\thanksref{8} \and
Liang Zhan\thanksref{8} \and
Yonghua Zhan\thanksref{16} \and
Aiqiang Zhang\thanksref{10} \and
Bin Zhang\thanksref{29} \and
Binting Zhang\thanksref{8} \and
Feiyang Zhang\thanksref{23} \and
Han Zhang\thanksref{8} \and
Hangchang Zhang\thanksref{8} \and
Haosen Zhang\thanksref{8} \and
Honghao Zhang\thanksref{16} \and
Jialiang Zhang\thanksref{21} \and
Jiawen Zhang\thanksref{8} \and
Jie Zhang\thanksref{8} \and
Jingbo Zhang\thanksref{17} \and
Junwei Zhang\thanksref{22} \and
Lei Zhang\thanksref{21} \and
Ping Zhang\thanksref{23} \and
Qingmin Zhang\thanksref{27} \and
Rongping Zhang\thanksref{8} \and
Shiqi Zhang\thanksref{16} \and
Shuihan Zhang\thanksref{8} \and
Siyuan Zhang\thanksref{22} \and
Tao Zhang\thanksref{23} \and
Xiaomei Zhang\thanksref{8} \and
Xin Zhang\thanksref{8} \and
Xu Zhang\thanksref{8} \and
Xuantong Zhang\thanksref{8} \and
Yibing Zhang\thanksref{8} \and
Yinhong Zhang\thanksref{8} \and
Yiyu Zhang\thanksref{8} \and
Yongpeng Zhang\thanksref{8} \and
Yu Zhang\thanksref{8} \and
Yuanyuan Zhang\thanksref{24} \and
Yumei Zhang\thanksref{16} \and
Zhenyu Zhang\thanksref{26} \and
Zhijian Zhang\thanksref{14} \and
Jie Zhao\thanksref{8} \and
Runze Zhao\thanksref{8} \and
Shujun Zhao\thanksref{29} \and
Tianhao Zhao\thanksref{8} \and
Hua Zheng\thanksref{14} \and
Yangheng Zheng\thanksref{11} \and
Li Zhou\thanksref{8} \and
Shun Zhou\thanksref{8} \and
Tong Zhou\thanksref{8} \and
Xiang Zhou\thanksref{26} \and
Xing Zhou\thanksref{8} \and
Jingsen Zhu\thanksref{16} \and
Kangfu Zhu\thanksref{27} \and
Kejun Zhu\thanksref{8} \and
Zhihang Zhu\thanksref{8} \and
Bo Zhuang\thanksref{8} \and
Honglin Zhuang\thanksref{8} \and
Liang Zong\thanksref{10} \and
Jiaheng Zou\thanksref{8}
}

\institute{
Yerevan Physics Institute, Yerevan, Armenia \label{1} \and
Universit\'{e} Libre de Bruxelles, Brussels, Belgium \label{2} \and
Universidade Estadual de Londrina, Londrina, Brazil \label{3} \and
Pontificia Universidade Catolica do Rio de Janeiro, Rio de Janeiro, Brazil \label{4} \and
Millennium Institute for SubAtomic Physics at the High-energy Frontier (SAPHIR), ANID, Chile \label{5} \and
Universidad Andres Bello, Fernandez Concha 700, Chile \label{6} \and
Beijing Institute of Spacecraft Environment Engineering, Beijing, China \label{7} \and
Institute of High Energy Physics, Beijing, China \label{8} \and
North China Electric Power University, Beijing, China \label{9} \and
Tsinghua University, Beijing, China \label{10} \and
University of Chinese Academy of Sciences, Beijing, China \label{11} \and
College of Electronic Science and Engineering, National University of Defense Technology, Changsha, China \label{12} \and
Chongqing University, Chongqing, China \label{13} \and
Dongguan University of Technology, Dongguan, China \label{14} \and
Jinan University, Guangzhou, China \label{15} \and
Sun Yat-Sen University, Guangzhou, China \label{16} \and
Harbin Institute of Technology, Harbin, China \label{17} \and
The Radiochemistry and Nuclear Chemistry Group in University of South China, Hengyang, China \label{18} \and
Wuyi University, Jiangmen, China \label{19} \and
Shandong University, Jinan, China, and Key Laboratory of Particle Physics and Particle Irradiation of Ministry of Education, Shandong University, Qingdao, China \label{20} \and
Nanjing University, Nanjing, China \label{21} \and
Guangxi University, Nanning, China \label{22} \and
School of Physics and Astronomy, Shanghai Jiao Tong University, Shanghai, China \label{23} \and
Tsung-Dao Lee Institute, Shanghai Jiao Tong University, Shanghai, China \label{24} \and
Nankai University, Tianjin, China\label{25} \and
Wuhan University, Wuhan, China\label{26} \and
Xi'an Jiaotong University, Xi'an, China\label{27} \and
Xiamen University, Xiamen, China\label{28} \and
School of Physics and Microelectronics, Zhengzhou University, Zhengzhou, China\label{29} \and
Institute of Physics, National Yang Ming Chiao Tung University, Hsinchu\label{30} \and
National United University, Miao-Li \label{31} \and
Department of Physics, National Taiwan University, Taipei \label{32} \and
Charles University, Faculty of Mathematics and Physics, Prague, Czech Republic \label{33} \and
University of Jyvaskyla, Department of Physics, Jyvaskyla, Finland \label{34} \and
IJCLab, Universit\'{e} Paris-Saclay, CNRS/IN2P3, 91405 Orsay, France \label{35} \and
Univ. Bordeaux, CNRS, LP2I, UMR 5797, F-33170 Gradignan, F-33170 Gradignan, France \label{36} \and
IPHC, Universit\'{e} de Strasbourg, CNRS/IN2P3, F-67037 Strasbourg, France \label{37} \and
Aix Marseille Univ, CNRS/IN2P3, CPPM, Marseille, France \label{38} \and
SUBATECH, Universit\'{e} de Nantes,  IMT Atlantique, CNRS-IN2P3, Nantes, France \label{39} \and
III. Physikalisches Institut B, RWTH Aachen University, Aachen, Germany \label{40} \and
Institute of Experimental Physics, University of Hamburg, Hamburg, Germany \label{41} \and
Forschungszentrum J\"{u}lich GmbH, Nuclear Physics Institute IKP-2, J\"{u}lich, Germany \label{42} \and
Institute of Physics and EC PRISMA$^+$, Johannes Gutenberg Universit\"{a}t Mainz, Mainz, Germany \label{43} \and
Technische Universit\"{a}t M\"{u}nchen, M\"{u}nchen, Germany \label{44} \and
Helmholtzzentrum f\"{u}r Schwerionenforschung, Planckstrasse 1, D-64291 Darmstadt, Germany \label{45} \and
Eberhard Karls Universit\"{a}t T\"{u}bingen, Physikalisches Institut, T\"{u}bingen, Germany \label{46} \and
INFN Catania and Dipartimento di Fisica e Astronomia dell Universit\`{a} di Catania, Catania, Italy \label{47} \and
Department of Physics and Earth Science, University of Ferrara and INFN Sezione di Ferrara, Ferrara, Italy \label{48} \and
INFN Sezione di Milano and Dipartimento di Fisica dell Universit\`{a} di Milano, Milano, Italy \label{49} \and
INFN Milano Bicocca and University of Milano Bicocca, Milano, Italy \label{50} \and
INFN Milano Bicocca and Politecnico of Milano, Milano, Italy \label{51} \and
INFN Sezione di Padova, Padova, Italy \label{52} \and
Dipartimento di Fisica e Astronomia dell'Universit\`{a} di Padova and INFN Sezione di Padova, Padova, Italy \label{53} \and
INFN Sezione di Perugia and Dipartimento di Chimica, Biologia e Biotecnologie dell'Universit\`{a} di Perugia, Perugia, Italy \label{54} \and
Laboratori Nazionali di Frascati dell'INFN, Roma, Italy \label{55} \and
University of Roma Tre and INFN Sezione Roma Tre, Roma, Italy \label{56} \and
Pakistan Institute of Nuclear Science and Technology, Islamabad, Pakistan \label{57} \and
Joint Institute for Nuclear Research, Dubna, Russia \label{58} \and
Institute for Nuclear Research of the Russian Academy of Sciences, Moscow, Russia \label{59} \and
Lomonosov Moscow State University, Moscow, Russia \label{60} \and
Comenius University Bratislava, Faculty of Mathematics, Physics and Informatics, Bratislava, Slovakia \label{61} \and
High Energy Physics Research Unit, Faculty of Science, Chulalongkorn University, Bangkok, Thailand \label{62} \and
National Astronomical Research Institute of Thailand, Chiang Mai, Thailand \label{63} \and
Suranaree University of Technology, Nakhon Ratchasima, Thailand \label{64} \and
The University of Liverpool, Department of Physics, Oliver Lodge Laboratory, Oxford Str., Liverpool L69 7ZE, UK, United Kingdom \label{65} \and
University of Warwick, University of Warwick, Coventry, CV4 7AL, United Kingdom \label{66} \and
Department of Physics and Astronomy, University of California, Irvine, California, USA \label{67} \and
\emph{Present Address:} Emirates Nuclear Technology Center (ENTC), Khalifa University, Abu Dhabi, United Arab Emirates \label{present}
}

\thankstext{email_jpc}{e-mail: \href{mailto:Juno_pub_comm@juno.ihep.ac.cn}{Juno\_pub\_comm@juno.ihep.ac.cn} (corresponding author)}

\date{Received: date / Accepted: date}

\maketitle

\twocolumn

\begin{abstract}
Large-scale organic liquid scintillator detectors \linebreak are highly efficient in the detection of MeV-scale electron antineutrinos. These signal events can be detected through inverse beta decay on protons, which produce a positron accompanied by a neutron. A noteworthy background for antineutrinos coming from nuclear power reactors and from the depths of the Earth (geoneutrinos) is generated by ($\alpha,\,n$) reactions. In organic liquid scintillator detectors, $\alpha$ particles emitted from intrinsic contaminants such as $^{238}$U, $^{232}$Th, and $^{210}$Pb/$^{210}$Po, can be captured on $^{13}$C nuclei, followed by the emission of a MeV-scale neutron. Three distinct interaction mechanisms can produce prompt energy depositions preceding the delayed neutron capture, leading to a pair of events correlated in space and time within the detector. Thus, ($\alpha,\,n$) reactions represent an indistinguishable background in liquid scintillator-based antineutrino detectors, where their expected rate and energy spectrum are typically evaluated via Monte Carlo simulations. This work presents results from the open-source SaG4n software, used to calculate the expected energy depositions from the neutron and any associated de-excitation products. Also simulated is a detailed detector response to these interactions, using a dedicated Geant4-based simulation software from the JUNO experiment. An expected measurable $^{13}$C$(\alpha,\,n)^{16}$O event rate and reconstructed prompt energy spectrum with associated uncertainties, are presented in the context of JUNO, however, the methods and results are applicable and relevant to other organic liquid scintillator neutrino detectors.
\end{abstract}

\section{Introduction}
\label{sec:intro}

Over the decades since the first experimental evidence of neutrino existence by Cowan and Reines in 1956~\cite{CowanReines}, liquid scintillator (LS) detectors have played a central role in neutrino physics. LS detectors of increasing size and improved performance have been developed, boasting broad physics programs. These detectors represent, so far, the only technology to detect reactor neutrinos at different baselines; in searches for sterile neutrinos (NEOS~\cite{NEOS}, STEREO \cite{STEREO}, PROSPECT~\cite{PROSPECT}, DANSS~\cite{DANSS}), measurement of neutrino oscillation parameters $\theta_{13}$ (Daya Bay~\cite{DAYABAY}, RENO \cite{RENO}, Double Chooz~\cite{DOUBLECHOOZ}), or the so-called {\it solar parameters} $\theta_{12}$ and $\Delta m^2_{21}$ (KamLAND~\cite{KAMLANDONOFF}). The same detection technique has been exploited to measure geoneutrinos, as demonstrated by KamLAND~\cite{KAMLANDGEO} and Borexino~\cite{BOREXINOGEO}. Outside of antineutrino detection, Borexino has provided world-leading measurements of solar neutrinos, thanks to its unprecedented radio-purity~\cite{BOREXINOpp}. SNO+ is also entering on the scene, with the primary goal to search for 0$\nu \beta \beta$ decay~\cite{SNOPLUS}, but also to measure reactor and geoneutrinos~\cite{SNO+LSREAC}. JUNO~\cite{JUNO_Yellow_Book, JUNOPHYSICS} is the first multi-kiloton LS detector, under construction in the South of China. Its design is driven by its main physics goal to determine the neutrino mass ordering~\cite{JUNONMO}, through precise measurement of the oscillation pattern in the energy spectrum of reactor neutrinos at a \SI{52.5}{km} baseline.

\begin{figure*}[t]
\centering
\includegraphics[width = 0.7\textwidth]{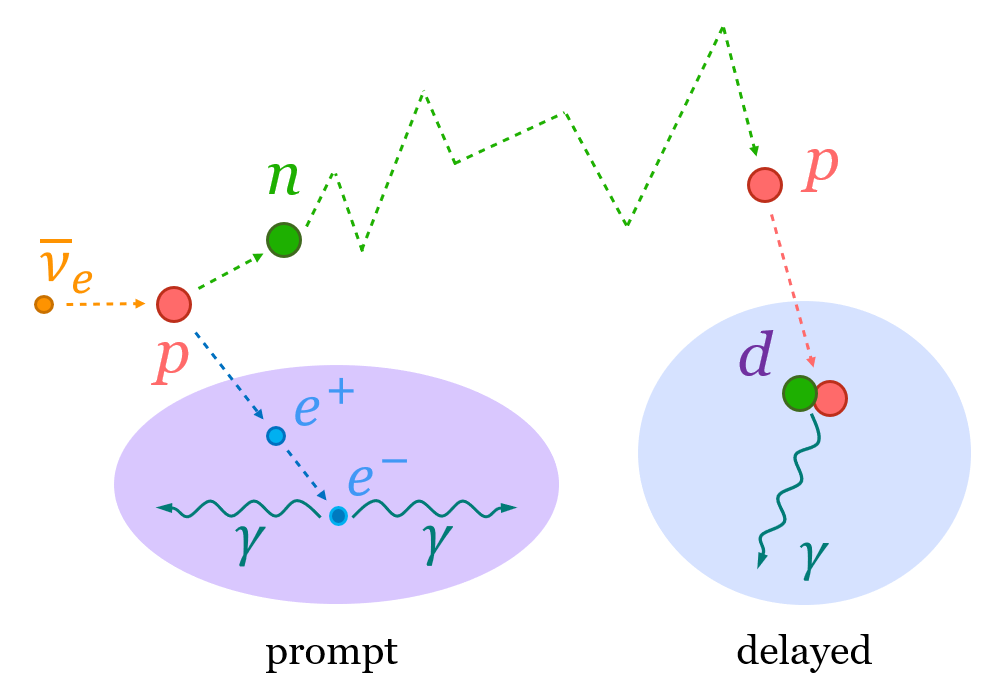}
\caption{Schematic of the IBD reaction on proton used for electron antineutrino detection in LS detectors.  It demonstrates the origin of the prompt
(violet area) and the delayed (blue area) signals, and underlines the similarity with the background caused by $(\alpha,\,n)$ reactions shown in figure~\protect\ref{fig:alphan_reac}.}
\label{fig:ibd_cartoon}
\end{figure*}

Detection of reactor electron antineutrinos is made \linebreak through the charged-current Inverse Beta Decay (IBD) reaction on protons:
\begin{equation}
\bar{\nu}_{e} + p \rightarrow e^{+} + n.
\label{eq:IBD}
\end{equation} 
IBD interactions feature a minimum antineutrino kinetic energy threshold of \SI{1.8}{MeV}, corresponding to the mass difference between the emitted particles, namely the neutron $n$ and positron $e^{+}$, and the initial proton $p$. The products of this reaction, schematized in figure~\ref{fig:ibd_cartoon}, yield a distinct coincident signal. The positron $e^{+}$ deposits its kinetic energy in the LS, then quickly annihilates with an electron in the detector, producing detectable scintillation light. This {\it prompt} signal bears information of the energy of the incident neutrino. The emitted neutron propagates on a random walk, quickly thermalizing via elastic collisions, typically with protons in the detector, until it is eventually captured by a proton/nucleus in the detector. Neutron capture on protons yields a deuteron followed by a \SI{2.2}{MeV} $\gamma$. Neutrons can also be captured on $^{12}$C in organic LS that leads to a \SI{4.95}{MeV} $\gamma$ emission. This option takes place with about 1\% probability.  The {\it delayed} event typically happens in LS with an averaged lifetime of roughly \SI{200}{\micro s}, where its precise value depends on the exact LS composition. Moreover, if LS is doped with gadolinium, neutron captures predominantly occur on isotopes of this chemical element, the capture time is substantially reduced, and a series of $\gamma$s with a total energy of \SI{8}{MeV} is emitted~\cite{SKGDLOAD}. In any case, the capture usually happens tens of centimeters away from the IBD interaction point. The distinct prompt-delayed space and time coincidence is a powerful characteristic for the selection of antineutrino interactions.

In spite of the background suppression power of the IBD coincidence tag, several background categories pose important challenges in antineutrino detection. Cosmogenic or accidental coincidence pair backgrounds, for example, can be evaluated and constrained in analysis by exploiting independent data sets. This can be done by collecting the events following cosmogenic muons and using off-time windows in the search for IBD-like events, respectively. 
Another correlated background, which can mimic the IBD signal, is known as the ($\alpha$,\,$n$) reaction, the focus of this work. In organic LS detectors, where there are large amounts of carbon, the dominant ($\alpha$,\,$n$) reaction occurs on $^{13}$C (with a natural abundance of 1.1\%~\cite{13C_abundance}). This produces $^{16}$O, often in an excited state, alongside a MeV-scale neutron. Preceding the delayed neutron capture, prompt signals can be generated by inelastic scattering of the neutron, along with higher energy radiation emitted upon the de-excitation of \textsuperscript{16}O, leading to correlated event pairs within the detector. Thus, the ($\alpha,\,n$) reactions represent an indistinguishable background in LS-based antineutrino detection. It is worth noting that ($\alpha$,\,$n$) reactions can also act as a background in direct searches for dark matter~\cite{an_white_paper}.

In general, an $\alpha$ can induce neutron emission from a nucleus in the LS cocktail via several reactions. The most basic one is the breakup of a deuteron into a proton and a neutron:
\begin{equation}
    \alpha + \textrm{$^2$H} \rightarrow \alpha + p + n.
    \label{eq:D_breakup}
\end{equation}
This process is significantly mitigated by the small natural abundance of deuterium ($0.001-0.028\%$~\cite{2H_abundance} versus 1.1\%~\cite{13C_abundance} for $^{13}$C), though the cross sections of \linebreak the $^{13}$C($\alpha$,\,$n$)$^{16}$O reaction and the deuteron breakup are comparable~\cite{Mohr18, JENDL, d_breakup_cross_section_1, d_breakup_cross_section_2}. Other reactions are related to the presence of additional chemical elements in the fluor and/or wavelength shifter, which in the case of JUNO are oxygen and nitrogen, as shown in section~\ref{sec:JUNO}. The respective neutron production rates are also negligible because of the small fractions of these components in the scintillator (from several mg/L to a few g/L) and the high energy thresholds of the processes, which are comparable to or exceed the maximum $\alpha$ energies in the $^{232}$Th and $^{238}$U decay chains ($\sim$\num{8}-\SI{9}{MeV}).

This work focuses on the evaluation of the ($\alpha$,\,$n$) background, the prediction of which strongly relies on Monte Carlo (MC) simulations and cannot be evaluated from independent datasets. In several experiments, the principal $\alpha$ particle source assumed to produce ($\alpha$,\,$n$) reactions, was \linebreak $^{210}$Po~\cite{BOREXINOGEO, KAMLANDONOFF,SNO+LSREAC}, but these reactions can also be sourced by $\alpha$ particles of various energies produced along the $^{238}$U and $^{232}$Th decay chains~\cite{DAYABAYOSC}. The relative contribution of different $\alpha$-producing isotopes depends on the achieved radiopurity of the LS. 

Presented here are results for the ($\alpha$,\,$n$) background simulated in the JUNO experiment. First introduced is the JUNO detector in section~\ref{sec:JUNO}. 
The assumed sources of $\alpha$s in LS are detailed in section~\ref{sec:AlphaSource}. The main characteristics of the \linebreak $^{13}$C($\alpha$,\,$n$)$^{16}$O reactions and the generation mechanisms of the prompt and delayed signals are then described in section~\ref{sec:alphanProc}. $^{13}$C($\alpha$,\,$n$)$^{16}$O reactions are simulated in the LS target using SaG4n v1.3 software~\cite{Men20, SaG4n_website}, presented in section~\ref{sec:genSag4n}, alongside the estimated interaction rates and neutron yields, with respective uncertainties. Products of the ($\alpha$,\,$n$) reactions, which deposit energy in the LS predicted by SaG4n, are then input to the JUNO simulation software \linebreak (JUNOSW)~\cite{JUNOSW}, implementing the full detector response and event reconstruction, which is covered in section~\ref{sec:genJunosw}. Also presented in this section is the selection procedure of IBD-like events due to $(\alpha,\,n)$ reactions, and the final expected measurable spectral shapes. Section~\ref{sec:event_rates} summarises the expected IBD-like background event rates due to $^{13}$C($\alpha$,\,$n$)$^{16}$O reactions, based on expected natural radioactivity concentrations in JUNO, along with discussion of the various systematic uncertainties.  Section~\ref{sec:conclusion} concludes the article and summarises the results, highlighting their possible applications and relevance to other organic LS-based neutrino detectors. 

\begin{figure*}[t]
\centering
\includegraphics[width = 0.95\textwidth]{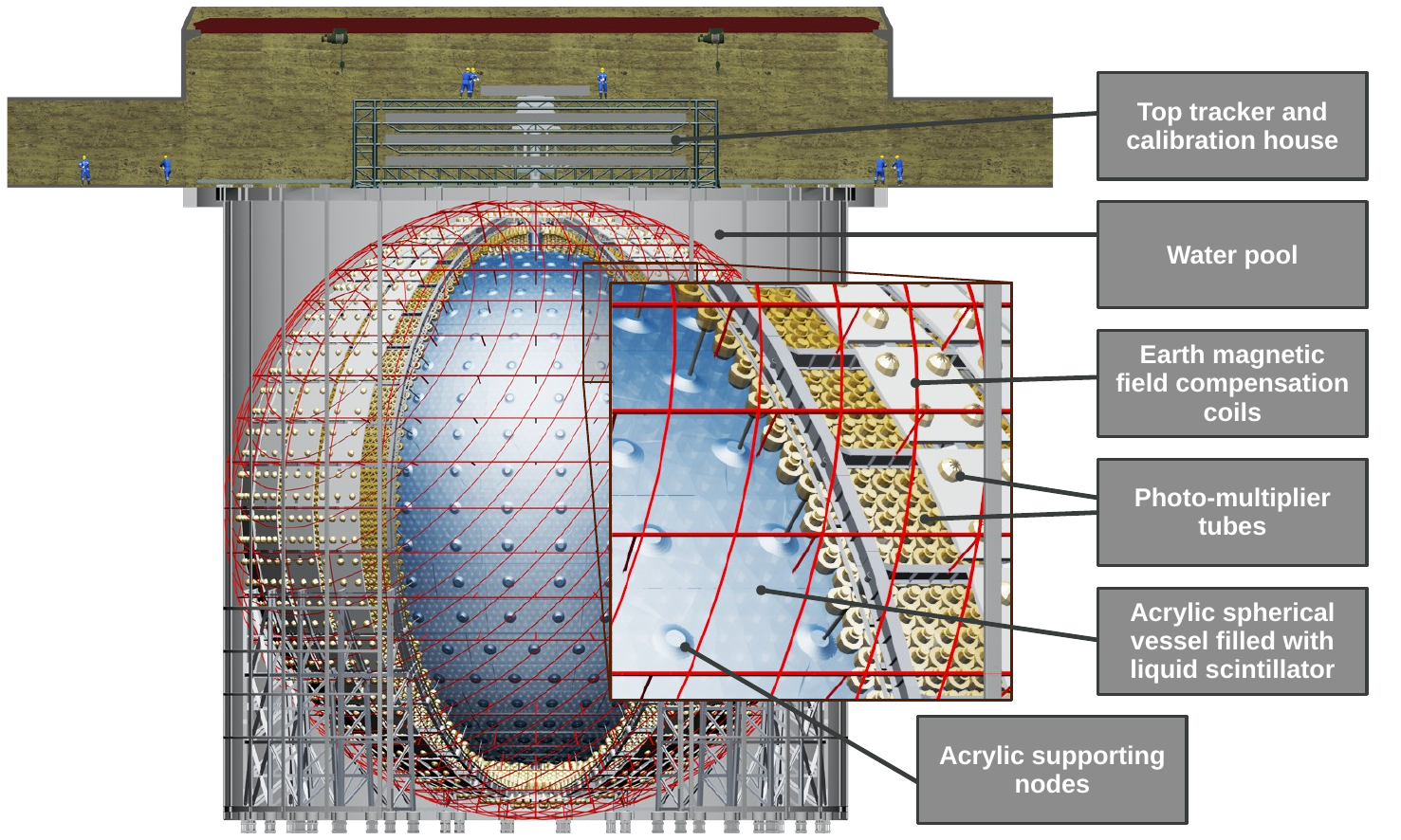}
\caption{Schematic drawing of the main JUNO detector.}
\label{fig:detector}
\end{figure*}

\section{The JUNO detector}
\label{sec:JUNO} 

The Jiangmen Underground Neutrino Observatory (JUNO) experiment is a 20-kiloton LS detector in southern China in an underground laboratory with a vertical overburden of $\approx$\SI{650}{m} (1800 m.w.e.). JUNO's primary physics goal is to measure the neutrino mass ordering (NMO), by resolving the fine structure due to flavor oscillations in the antineutrino energy spectrum from nearby nuclear reactors. In order to achieve this precision measurement, the detector is expected to reach an unprecedented energy resolution of 3$\%$ at \SI{1}{MeV}~\cite{JUNOEnergyResPaper2024}. 

A sketch of the JUNO detector is shown in figure~\ref{fig:detector}. It consists of a Central Detector (CD), containing \SI{20}{kt} of LS in a \SI{17.7}{m} radius acrylic sphere of \SI{120}{mm} thickness. The acrylic vessel is supported by a spherical stainless steel (SS) structure via 590 connecting bars. The LS target is watched by 17,612 20-inch and 25,600 3-inch photomultiplier tubes (PMTs) mounted on the SS structure. This allows for a first-rate photosensitive coverage (75.2\% for the 20-inch and 2.7\% for the 3-inch PMTs), which is needed to collect a large number of photoelectrons per unit of deposited energy in the scintillator. 

\begin{table*}[t]
\centering
\caption{Summary of $\alpha$ decaying isotopes from the $^{238}$U and $^{232}$Th chains in secular equilibrium, showing the respective half-lives $\tau_{1/2}$, \linebreak $\alpha$ energies $E_{\alpha}$, and branching ratios $BR_{\alpha}$ based on NuDAT~\protect\cite{NuDAT}. Branches with $BR_{\alpha}$ less than 1\% are not shown here but considered in \linebreak the analysis and depicted in lower part of figure~\protect\ref{fig:xs_and_a_energies}.}
\label{tab:alpha_source} 
\vskip 2pt
\begin{tabular}
{c |  c  c  c  | c | c c c} 
\hline
\hline
\multicolumn{4}{c|}{$^{238}$U chain}       &\multicolumn{4}{c}{$^{232}$Th chain}  \\
\multicolumn{1}{c}{Parent}&  $\tau_{1/2}$ &  $E_{\alpha}$ [MeV]   & $BR_{\alpha}$ [\%]& \multicolumn{1}{c}{Parent}&  $\tau_{1/2}$ &  $E_{\alpha}$ [MeV]   & $BR_{\alpha}$ [\%]    \\%
\hline
$^{238}$U       & \SI{4.4E9}{y}       & 4.198   & 79.0          &$^{232}$Th      & \SI{1.4E10}{y}      & 4.012   & 78.2  \\
                &                     & 4.151   & 20.9          &                &                     & 3.947   & 21.7 \\  
$^{234}$U       & \SI{2.4E5}{y}       & 4.774   & 71.38         &$^{228}$Th      & \SI{1.91}{y}        & 5.423   & 73.4 \\
                &                     & 4.722   & 28.42         &                &                     & 5.340   & 26.0   \\ 
$^{230}$Th      & \SI{7.5E4}{y}       & 4.687   & 76.3          &$^{224}$Ra      & \SI{3.66}{\day}     & 5.685   & 94.92 \\
                &                     & 4.620   & 23.4          &                &                     & 5.448   & 5.06 \\   
$^{226}$Ra      & \SI{1600}{y}        & 4.784   & 93.84         &$^{220}$Rn      & \SI{55}{\second}    & 6.288   & 99.88  \\
                &                     & 4.601   & 6.16          &$^{216}$Po      & \SI{0.14}{\second}  & 6.778   & 99.99\\                 
$^{222}$Rn      & \SI{3.82}{\day}     & 5.489   & 99.92         &$^{212}$Bi      & \SI{61}{\minute}    & 6.089   & 9.74 \\
$^{218}$Po      & \SI{3.098}{\minute} & 6.002   & 99.98         &                &                     & 6.050   & 25.12 \\
$^{214}$Po      & \SI{164.3}{\micro\second} & 7.686   & 99.96   &                & via $^{212}$Po      & 8.784   & 64.06\\
$^{210}$Po      & \SI{138.3}{\day}    & 5.304   & 99.99         &                & (\SI{3E-7}{\second})&         &   \\

\hline
\hline
\end{tabular}
\end{table*}   

The LS cocktail has been optimized in dedicated studies with the Daya Bay detector~\cite{DBJUNOLS}. The LS is primarily made up of linear alkylbenzene (LAB), consisting of long alkyl chains and typically containing 10-16 C atoms with a benzene ring attached at the end. JUNO employs a primary fluor in the form of 2,5-Diphenyloxazole (PPO), at a concentration of \SI{2.5}{g/L}, to avoid scintillation light re-absorption during its propagation within the detector. To increase the scintillation detection efficiency on PMTs, a wavelength shifter of p-bis-(o-methylstyryl)-benzene (bis-MSB) is also added at \SI{3}{mg/L}. LAB and its associated fluors were selected due to its high light yield, good $\alpha$/$\beta$ particle discrimination~\cite{Rebber2021}, and the ability to reach very high levels of purity. This scintillation mixture expects to allow for light attenuation lengths greater than \SI{20}{m} at \SI{430}{nm} in order to make up for the huge CD dimensions. In order to boost light collection, and reduce the levels of aforementioned naturally occurring radioactivity within the LS, it is passed through optical and radiochemical purification~\cite{JUNOPURIF}. A pre-detector (OSIRIS) also monitors 15\% of the LS for its radioactive contamination levels prior to filling into the JUNO detector~\cite{JUNOOSIRIS}. 

The CD is submerged in a cylindrical water pool (WP) of \SI{43.5}{m} diameter and height of \SI{44.0}{m}, filled with \SI{35}{kt} of ultrapure water. The WP shields the CD against external fast neutrons and $\gamma$s. It also acts as a Cherenkov veto for atmospheric muons, which have a flux of about \linebreak \SI{4E-3}{m^{-2}s^{-1}}. Cherenkov light produced by muons passing through water can be detected by the 2,400 20-inch PMTs installed on the outer surface of SS structure. On the top of the WP, a Top Tracker (TT) is placed to precisely measure the tracks of a subsample of the crossing muons~\cite{JUNOTT}.

Multiple calibration systems implementing radioactive and laser-based sources have been developed to calibrate the detector and to evaluate the non-uniformity and non-linearity of its response. The employed radioactive sources include $\gamma$ sources of various energies, a $^{68}$Ge positron source, \ce{^{241}Am-Be} (AmBe) and \ce{^{241}Am-^{13}C} (AmC) neutron sour\-ces. Calibration operations will be carried out through a stainless steel chimney, which connects the CD to the outside from the top. Calibration sources can be deployed throughout the inside of the acrylic vessel using an Automatic Calibration Unit (ACU)~\cite{ACU}, which covers the central axis, while the Cable Loop System (CLS)~\cite{CLS} allows for coverage of the off-axis region in a two-dimensional plane. A Guide Tube Calibration System (GTCS)~\cite{GTCS} can place sources on the outer surface of the acrylic sphere. Details regarding the calibration systems and strategies can be found in~\cite{JUNOCALIB}.

JUNO’s world-leading size, low backgrounds, and unprecedented energy resolution, allow for a very broad \linebreak physics program, measuring neutrinos from various sources, ranging in energy from tens of keV to tens of GeV~\cite{JUNO_Yellow_Book, JUNOPHYSICS}. Beyond reactor antineutrinos, JUNO will be able to detect solar, geo, atmospheric, and supernovae neutrinos, and to search for evidences of physics beyond the Standard Model (BSM)~\cite{JUNO8B,JUNO7BECNO,JUNOCCSN,JUNODIFFUSE,JUNOATMOS,JUNOND,JUNOPHYSICS}.

\section{Sources of \texorpdfstring{$\alpha$}{alpha} particles}
\label{sec:AlphaSource}

Liquid scintillators employed in neutrino detectors \linebreak undergo complex purification procedures, strongly reducing its radioactivity. Nevertheless, residual impurities do contain $\alpha$ emitting isotopes triggering $^{13}$C$(\alpha,\,n)^{16}$O reaction. \linebreak The most common source of $\alpha$s was found to be \linebreak $^{210}$Po~\cite{BOREXINOGEO, KAMLANDONOFF, SNO+LSREAC}, the last radioactive isotope of $^{238}$U chain, often breaking the secular equilibrium of the chain and contaminating the LS in much increased levels. Out-of-equilib\-rium $^{210}$Po with half-life of 138.4 days can be brought to LS stand-alone from external materials due to its chemical properties and mobility~\cite{BOREXINOCNO}. $^{238}$U chain secular equilibrium is often broken also by increased levels of relatively longed lived $^{210}$Pb. With its 22-year half-life, $^{210}$Pb represents a steady source of $^{210}$Po in the LS, via 
\begin{align}
^{210}\mathrm{Pb}\,(\beta^{-}, Q = \SI{63.5}{keV}) &\rightarrow  \,
^{210}\mathrm{Bi}\,(\beta ^{-}, Q = \SI{1.16}{MeV}) \rightarrow  \, \\ \nonumber
&\rightarrow  \,
^{210}\mathrm{Po}\,(\alpha, Q = \SI{5.407}{MeV}).
\label{eq:210Pb_decay}
\end{align}
Moreover, $^{238}$U chain also includes another long-lived nuclide, namely $^{226}$Ra with 1600-year half-life, which is a source of $^{222}$Rn and a series of short-lived daughters. The respective out-of-equilibrium component~\cite{an_white_paper} might be present in the JUNO CD, if, for example, there is an air leak, usually containing a large amount of $^{222}$Rn. This extra source of $\alpha$ particles can be easily identified by monitoring the rate of so-called Bi-Po events (see Sec.~\ref{sec:bipo}).

\begin{figure*}[t]
\centering
\includegraphics[width = 1.0\textwidth]{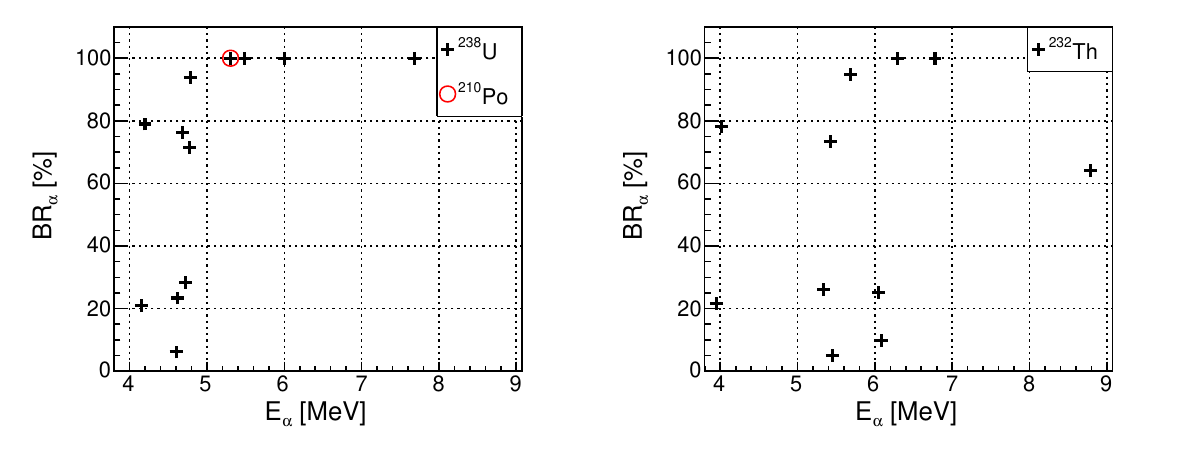}
\caption{The branching ratios of $\alpha$ particles from the $^{238}$U (left) and $^{232}$Th (right) radioactive chains in secular equilibrium (see also table~\protect\ref{tab:alpha_source}) are shown as a function of $\alpha$ energy. The $\alpha$ from $^{210}$Po, often breaking the equilibrium of the $^{238}$U chain, is marked with a red circle in the left figure. 
}
\label{fig:sag4n_alphaE}
\end{figure*}

Liquid scintillators typically contain residual amounts of $^{238}$U and $^{232}$Th in secular equilibrium, in which decays from all the daughter isotopes occur at the same rate. The $\alpha$ decaying isotopes, 8 from the $^{238}$U and 6 from the $^{232}$Th chains, produce, respectively, 12 and 11 $\alpha$s, as summarized in table~\ref{tab:alpha_source}, showing the respective half-lives, $\alpha$ energies, and branching ratios from NuDAT~\cite{NuDAT}. The relative weights of different $\alpha$s as a function of their energy in both chains are  visualized in figure~\ref{fig:sag4n_alphaE}.

The world's best LS radiopurity was achieved by Borexino~\cite{BorexinoBGs,BorexinoPhase2}, suppressing $^{238}$U and $^{232}$Th by ten orders of magnitude (less than $9.4 \times 10^{-20}$ g/g of $^{238}$U and less than $5.7 \times 10^{-19}$ g/g of $^{232}$Th at 95\% C.L.). This level of radiopurity was fundamental for the successful detection of solar neutrinos via elastic scattering off electrons. In Borexino, out-of-equilibrium $^{210}$Po was thus the only source of the overall almost negligible $(\alpha,\,n)$ background in the measurement of geoneutrinos~\cite{BOREXINOGEO}. Thanks to the IBD coincidence tag, experiments which focus on antineutrino detection do not require such extreme radiopurity levels. Nevertheless, the $(\alpha,\,n)$ background played an important role in KamLAND's reactor~\cite{KAMLANDREA} and geoneutrino measurements~\cite{KAMLANDGEO}, especially in its first phase before additional LS  purification. Recent reactor antineutrino measurements in SNO+ featured significant rates of $^{210}$Po-sourced ($\alpha,\,n$), which proved to be the most significant background~\cite{SNO+LSREAC}. The Daya Bay experiment considered $\alpha$ decays from $^{210}$Po, \textsuperscript{238}U, \textsuperscript{232}Th, and \textsuperscript{227}Ac, however, due to the very high reactor signal flux, ($\alpha,\,n$) was evaluated to occur at a negligible rate~\cite{DAYABAYOSC}. 

In this work, we evaluate the $^{13}$C($\alpha,\,n$)$^{16}$O background from $^{238}$U and $^{232}$Th chains, from out-of-equilibrium \linebreak $^{210}$Pb/$^{210}$Po, and from unsupported $^{210}$Po in JUNO. As the final LS radiopurity is not yet known, we consider the minimum radiopurity level requested for the NMO measurement, that is, $10^{-15}$ g/g for $^{238}$U and $^{232}$Th. $^{210}$Pb, which subsequently decays to $^{210}$Po, can fall out of equilibrium from the $^{238}$U chain, and is evaluated to be $5 \times 10^{-23}$ g/g relying on JUNO's radioactivity control strategy~\cite{Juno21, JUNO7BECNO}. The contamination from unsupported $^{210}$Po is $3 \times 10^{-22}$ g/g, based on a $^{210}$Po rate of \SI{8E4}{cpd/kt} (``cpd'' stands for counts per day) reported in Borexino as the average value in the whole LS volume at the beginning of data taking~\cite{BOREXINOGEO}. It is reasonable to assume that this initial contamination originated from the inner surfaces of the LS filling system and the target vessel. We assume the same contamination level of the surfaces in JUNO as in Borexino, accounting for differences in surface areas and LS volumes. The \textsuperscript{227}Ac $\alpha$ source, observed in the Daya Bay measurement, is not expected to have significant presence in JUNO and is therefore not considered in this work. The \textsuperscript{235}U decay chain, of which \textsuperscript{227}Ac is a daughter isotope, has a natural abundance of less than 1$\%$. The heightened concentration of \textsuperscript{227}Ac seen in Daya Bay was determined to originate from the Gd loaded in their LS, which will not be added to the JUNO LS cocktail during the NMO-measurement phase.

\section{\texorpdfstring{$^{13}$C$(\alpha,\,n)^{16}$O}{13C(alpha, n)16O} reaction in liquid scintillator}
\label{sec:alphanProc}

\begin{figure*}[t]
    \centering
    \includegraphics[width = \textwidth]{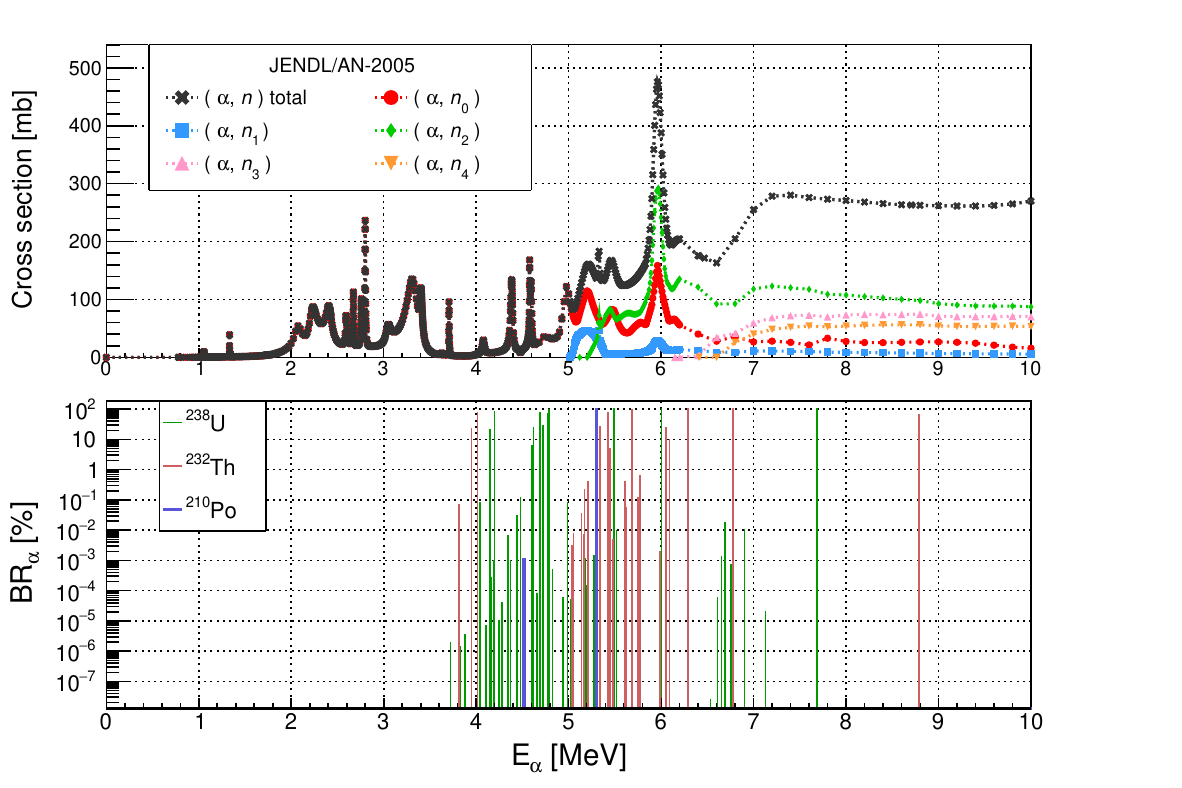}
    \caption{Top: Values of the $^{13}$C($\alpha$,\,$n$)$^{16}$O reaction cross-section as a function of the initial $\alpha$ energy are taken from the JENDL/AN-2005 data library~\protect\cite{JENDL} and shown with different markers for the considered excitation levels of oxygen. The total cross section is shown in black; note that below \SI{5}{MeV} this overlaps with red markers of the cross section for the case when $^{16}$O is created in the ground state $n_0$. The other colours mark the partial cross sections of the cases when $^{16}$O is produced in excited states up to $n_4$. 
     Dotted lines connecting data points are to only guide the eye.
   Bottom: The complete $\alpha$ spectra from $^{238}$U and $^{232}$Th chains assuming secular equilibrium and from $^{210}$Po decay, used in this study.}
    \label{fig:xs_and_a_energies}
\end{figure*}

The cross-section which quantifies the probability of \linebreak a $^{13}$C($\alpha$,\,$n$)$^{16}$O reaction occurring for a given incident $\alpha$ particle energy, used in this work, is shown in the top part of figure~\ref{fig:xs_and_a_energies}. These data are adopted from the JENDL/AN-2005 data library~\cite{JENDL}, implemented in SaG4n package as the only available evaluation of the $(\alpha,\,n)$ reactions cross-sections, which was calculated based on experimental data. The data points in the top plot of figure~\ref{fig:xs_and_a_energies} show multiple resonances, which is expected for ($\alpha$,\,$n$) reactions on light nuclides, such as $^{13}$C. This dependence is due to the complex mechanism of formation of a compound nucleus, which has numerous energy levels for possible excited states. The lower part of figure~\ref{fig:xs_and_a_energies} shows the complete $\alpha$ spectra from $^{238}$U and $^{232}$Th chains, including branches with $BR_{\alpha}$ below 1\%, that were not explicitly discussed in the previous section, but considered in the analysis. It can be seen that $\alpha$ particle energies extend from around \SI{3.5}{MeV} to \SI{9}{MeV}.

There are three distinct mechanisms by which \linebreak the $^{13}$C($\alpha$,\,$n$)$^{16}$O reaction can mimic an IBD coincidence pair, schematized and labeled in figure \ref{fig:alphan_reac}.  In each case, a neutron is emitted, producing an identical \textit{delayed} neutron capture event. The basic scenarios of the prompt formation can be described as follows:

\begin{figure*}[ht]
    \centering
    \includegraphics[width = 0.76\textwidth]{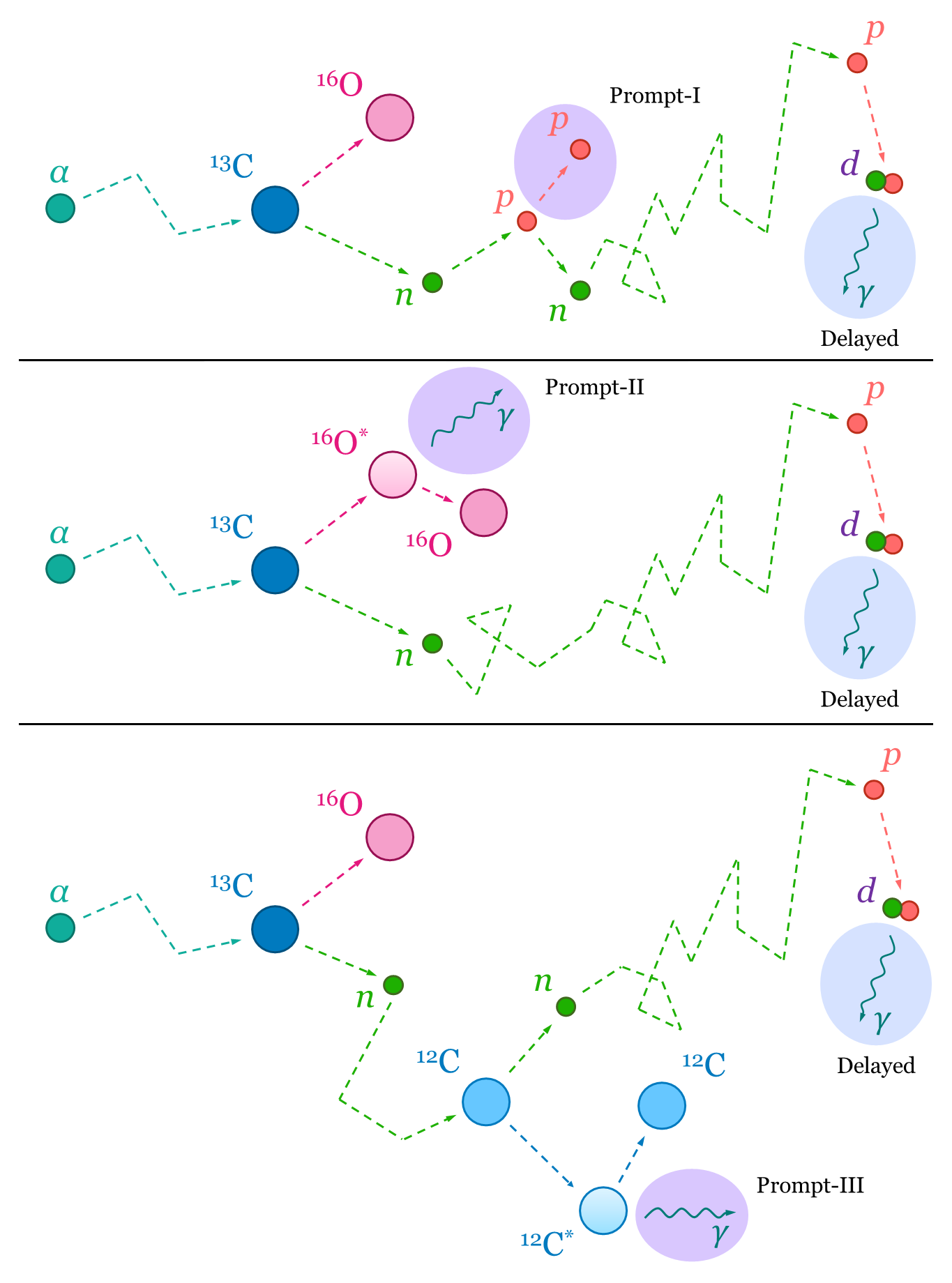}
    \caption{Scheme of the ($\alpha$,\,$n$) reaction on $^{13}$C. The three
processes that can generate the three kinds of the prompt signals, Prompt-I, II, and III, as described in text, are shown in violet areas. The blue area indicates the delayed signal from the neutron capture. Note that combinations of Prompt-I with Prompt-II or Prompt-III are also possible depending on the $\alpha$ energy, as discussed in the text.
}
    \label{fig:alphan_reac}
\end{figure*}

\begin{figure*}[t]
    \centering
    \includegraphics[width = 0.9\textwidth]{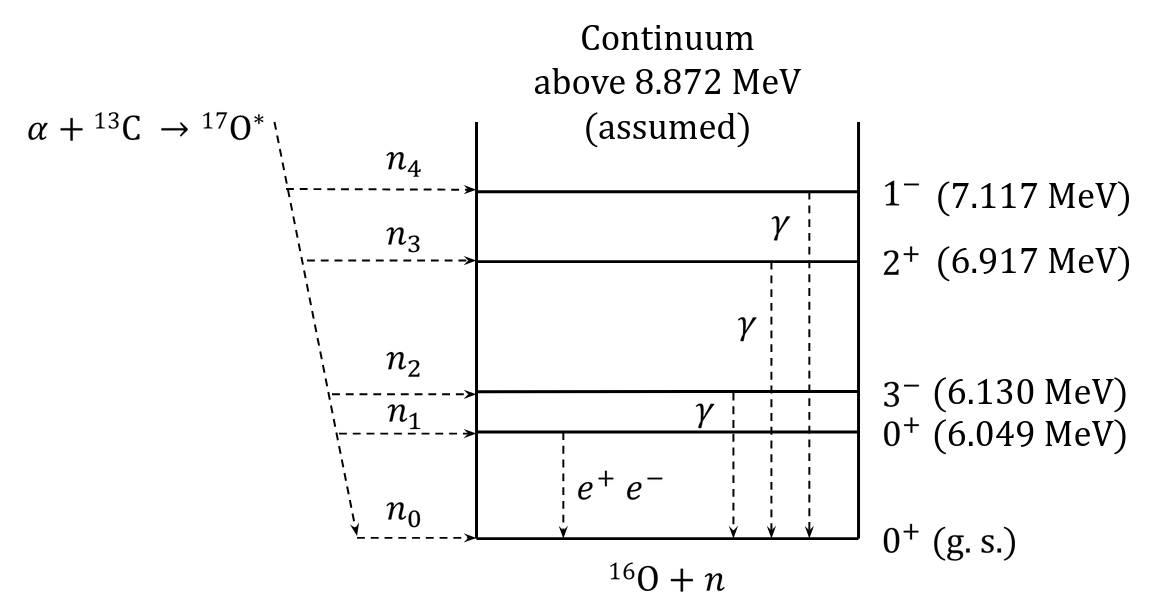}
    \caption{The simplified level scheme of $^{16}$O as populated in the $^{13}$C($\alpha$,\,$n$)$^{16}$O reaction~\protect\cite{NuDAT}. Values $n_0$, $n_1$, $n_2$, $n_3$, $n_4$ represent which final state in $^{16}$O is populated. }
    \label{fig:excited_states}
\end{figure*}

1.  {\bf Prompt-I from scattered protons}: The emitted neutron elastically scatters multiple protons within the first \linebreak $\mathcal{O}$\SI{}{(ns)} of its random walk, producing scintillation light. 

2.  {\bf Prompt-II from \textsuperscript{16}O$^{*}$ de-excitation}: Upon the capture of an $\alpha$ particle with a kinetic energy above $\sim$\SI{5}{MeV}, \textsuperscript{16}O may be produced in an excited state. For the first excited state,  $n_1$, during de-excitation, a pair of $e^+$ + $e^-$ is emitted with a total kinetic energy of 5.03 MeV. The annihilation of the positron with an electron in the detector yields $\gamma$s of total energy \SI{1.02}{MeV}, resulting in a prompt event with a total deposited energy of \SI{6.05}{MeV}. In the cases \textsuperscript{16}O is produced in its 2\textsuperscript{nd}, 3\textsuperscript{rd}, and 4\textsuperscript{th} excited states, $n_{2}$, $n_{3}$, and $n_{4}$, \linebreak transitions to the ground state release $\gamma$s with energies \linebreak \SI{6.130}{MeV}, \SI{6.917}{MeV}, and \SI{7.117}{MeV}, respectively, \linebreak as seen in figure \ref{fig:excited_states}. 

3. {\bf Prompt-III from \textsuperscript{12}C$^{*}$ de-excitation}:\textsuperscript{16}O is produced in its ground state, but the high energy neutron inelastically scatters off a \textsuperscript{12}C nucleus, prompting its excitation and subsequent emission of a \SI{4.4}{MeV} $\gamma$.

We note that the proton scattering, described in Prompt-I, also occurs in coincidence with the Prompt-II and Prompt-III mechanisms. In these cases, however, the available energy for the neutron has already been used in the excitation of either \textsuperscript{16}O or \textsuperscript{12}C, where proton scattering causes emission of only a small amount of scintillation light.

We also note that the $\alpha$ particle deposits a fraction of its initial kinetic energy into the LS before its capture on \textsuperscript{13}C. Furthermore, the quenching effect strongly decreases the visible energy produced by $\alpha$s, typically by an order of magnitude compared to $e^{+/-}$ and $\gamma$s. Consequently, the $\alpha$s yield small but measurable scintillation light which combines with each of the prompt processes described above. The quenching effect is discussed in more detail in \linebreak section~\ref{sec:quneching}.

Overall, $^{13}$C$(\alpha,\,n)^{16}$O reaction can produce IBD-like coincident signals with prompt energies up to about \SI{7}{MeV}, featuring a complex energy spectrum. The following section describes the simulation of $^{13}$C$(\alpha,\,n)^{16}$O reactions in LS using the SaG4n tool.
   
\section{\texorpdfstring{Simulation of the $^{13}$C$(\alpha,\,n)^{16}$O}{13C(alpha, n)16O} reaction in liquid scintillator}
\label{sec:genSag4n}

The first step in the evaluation of an IBD-like background from $^{13}$C$(\alpha,\,n)^{16}$O reaction is its simulation using \linebreak the Geant4-based simulation tool SaG4n~\cite{Men20, SaG4n_website}. SaG4n package version 1.3 with Geant4.11.1.2~\cite{Geant4-2003, Geant4-2006, Geant4-2016} was used in this work. Taking into account the incident $\alpha$s of different energies from expected radio-impurities, we simulate the energy loss by $\alpha$s in the LS until its eventual capture on $^{13}$C nuclei, accounting for the cross-sectional energy dependence. Several cross-section libraries are available within the program. We adopted the JENDLTENDL01 dataset, since it contains the aforementioned JENDL/AN-2005 cross section evaluations for capture on carbon. In section~\ref{subsec:sag4ninput} we describe our inputs and settings used in the SaG4n software, which can in general be used to simulate various $(\alpha,\,n)$ reactions in different materials. Section~\ref{subsec:sag4n_result} describes simulation results in terms of the branching ratios of different energy levels of the produced $^{16}$O nucleus, neutron yields, and neutron energy spectra. In section~\ref{subsec:system} we discuss various systematic effects that can impact our conclusions.

\subsection{SaG4n software settings and inputs}
\label{subsec:sag4ninput}

SaG4n requires definition of the target  material composition and geometry, sources of $\alpha$s, and of several parameters characterizing the simulation process. We simulated the $(\alpha,\,n)$ reaction as well as the de-excitation of the $^{16}$O nucleus, while all secondary particles are disabled. In this work, we have used the following definitions:

\begin{itemize}

\item ALPHA SOURCES: We simulated 2$\times10^9$ decays of \linebreak $^{210}$Po and the same number of alphas from $^{238}$U and $^{232}$Th chains in secular equilibrium. For the latter two, we used built-in SaG4n functions to provide the energies and relative intensities of each $\alpha$ decay within these chains, as seen in the lower part of figure~\ref{fig:xs_and_a_energies}. All $\alpha$s are emitted isotropically within a cube of \SI{10}{cm} side length placed in the center of the simulated target.

\item TARGET GEOMETRY: A cube of \SI{100}{cm} side length, sufficiently large with respect to the size of the $\alpha$ sources and mm range of $\alpha$s, guaranteed full energy deposition in the target. 

\item TARGET MATERIAL: JUNO LS was characterized \linebreak with a simplified chemical formula C$_6$H$_5$C$_{14}$H$_{29}$ and \linebreak with a density of \SI{0.853}{g/cm^3} at 20$^{\circ}$C. The corresponding mass fractions of hydrogen and carbon are 12.49\% and 87.51\%, respectively. Since the scintillator cocktail consist of 99.7\% LAB by mass, there is assumed negligible impact of the C nuclei present from the PPO and bis-MSB fluors. A natural abundance of $^{13}$C equal to $\sim$1.1\%~\cite{13C_abundance} was considered.

\item SIMULATION PARAMETERS: 
The SaG4n parameter named the {\it bias factor} allows one to magnify the $\alpha$ capture cross section in the material in order to reduce the computing time for the simulation of a desired number of events. Consequently, reaction products are generated with weights $\omega$, which take into account the enhancement of the ($\alpha,\,n$) cross section. A bias factor of 10$^4$ was assumed in this work. It was found that there was negligible impact to results when bias factors of 10$^5$ or 10$^6$ were used. Another important parameter is the {\it maximum allowed step length ($S_{\mathrm{max}}$)} for the propagation of $\alpha$ particles within the material. The chosen step length, unless stated otherwise, was \SI{1}{\micro\meter}, a factor of 10 smaller than SaG4n's default value. This choice is discussed in section~\ref{subsec:system}.
\end{itemize}

\subsection{Reaction products} 
\label{subsec:sag4n_result}

SaG4n outputs information about all energy-depositing particles involved in the ($\alpha,\,n$) reaction. For each simulated interaction, we recorded the energy, position, and direction of $\alpha$ at emission and capture on \textsuperscript{13}C as well as of the neutron and \textsuperscript{16}O de-excitation product(s).

The neutron yield $Y\lbrack n/\alpha \rbrack$ in SaG4n simulation, {\it i.e.} the probability per $\alpha$ to trigger a $^{13}$C$(\alpha,\,n)^{16}$O reaction, can be estimated as: 
\begin{equation}
Y\lbrack n/\alpha \rbrack = \frac{1}{N_{\alpha}} \sum_{i=1}^{N_n} \omega_{i},  
\end{equation} 
where $N_n$ is the total number of neutrons produced in simulation, $\omega_{i}$ is the weight of each neutron event, and $N_{\alpha}$ is the number of simulated initial $\alpha$ particles. For radioactive chains, we define neutron yield of the whole chain in secular equilibrium, $Y\,\lbrack n/\text{chain} \rbrack$, that is obtained by multiplying the $Y\lbrack n/\alpha \rbrack$ by the number of $\alpha$-decaying isotopes in each chain, e.g. 8 for $^{238}$U and 6 for $^{232}$Th.

\begin{figure*}[t]
\centering
\includegraphics[width = 0.95\textwidth]{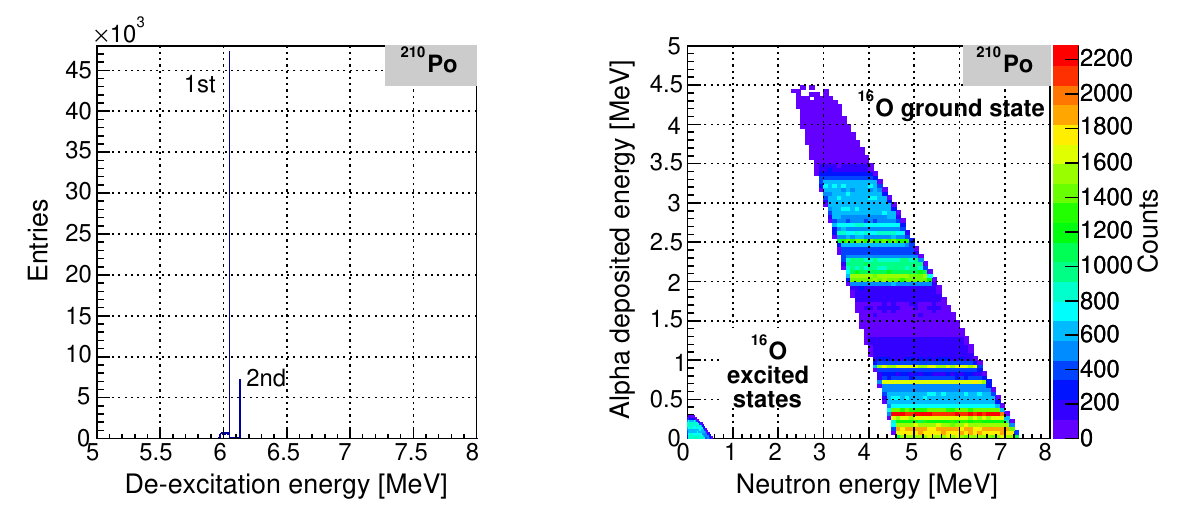}
\caption{SaG4n simulation of the $^{13}$C$(\alpha,\,n)^{16}$O reaction in the JUNO LS triggered by \SI{5.3}{MeV} $\alpha$ from $^{210}$Po. Left: spectrum of $e^+e^-$ pairs and $\gamma$s from de-excitation of the first and second \textsuperscript{16}O excited states, respectively. Right: correlation between the energy of emitted neutron and deposited energy of the $\alpha$ particle prior to its capture.
}
\label{fig:sag4n_pb210out}
\end{figure*}

Results from the simulation of $^{210}$Po are shown in figure~\ref{fig:sag4n_pb210out}. The \SI{5.3}{MeV} $\alpha$ allows population of not only the ground state but also of the 1\textsuperscript{st} and 2\textsuperscript{nd} excited states of \textsuperscript{16}O. The energy spectrum of de-excitation  $e^+e^-$ pairs\footnote{SaG4n actually generates a single $\gamma$ for the $e^+e^-$ mode, with a corresponding total energy of \SI{6.049}{MeV}. This technical feature is corrected in the next simulation stage, as described in section~\ref{subsec:sag4n_junosw}.} at \SI{6.049}{MeV} and $\gamma$s at \SI{6.130}{MeV} can be seen in the left plot of figure~\ref{fig:sag4n_pb210out}. The right part of this figure demonstrates the correlation between the energy of emitted neutron and the deposited energy of $\alpha$ particle prior to its capture. When \textsuperscript{16}O is produced in its ground state,
the emitted neutron acquires energies in the range of 3 to \SI{7}{MeV} and energy depositions from the $\alpha$ can extend up to about \SI{4.5}{MeV}. The horizontal bands clearly visible in the figure correspond to the fine structure in the $\alpha$ capture cross section, as shown in the top part of figure~\ref{fig:xs_and_a_energies}. Thus, as the $\alpha$ decreases in energy, the probability of its capture can increase by a factor of more than 100 at certain energies.  When \textsuperscript{16}O is produced in an excited state, most of the $\alpha$ energy is absorbed in the excitation itself. In these cases, the $\alpha$ deposits only a small amount of energy before its capture (well below \SI{0.5}{MeV}) and only similar amounts of energy are transferred to the emitted neutron. 

\begin{figure*}[htp]
\centering
\includegraphics[width = 0.90\textwidth]{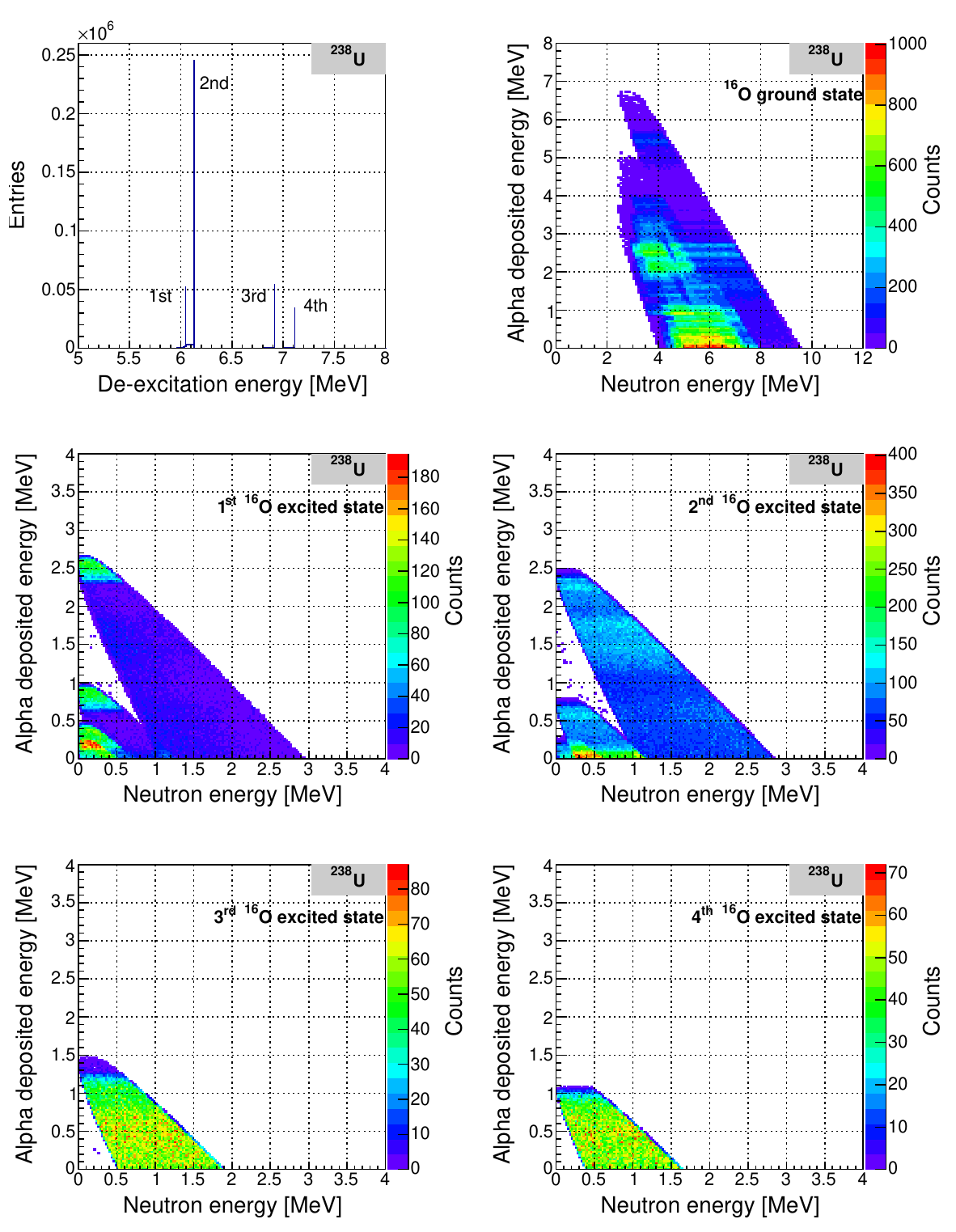}
\caption{SaG4n simulation of the $^{13}$C$(\alpha,\,n)^{16}$O reaction in the JUNO LS triggered by $\alpha$s from the $^{238}$U chain in secular equilibrium. Top left: spectrum of $e^+e^-$ pairs and $\gamma$s from de-excitation of the 1$^{\mathrm{st}}$ and the 2$^{\mathrm{nd}}$ to  4$^{\mathrm{th}}$ \textsuperscript{16}O excited states, respectively. Other plots show correlations between the energy of emitted neutron and the deposited energy of $\alpha$ particle prior to its capture: top right plot for the case  when \textsuperscript{16}O is created in its ground state, while the remaining plots for the cases of the 1$^{\mathrm{st}}$ to  4$^{\mathrm{th}}$ \textsuperscript{16}O excited states, respectively.}
\label{fig:sag4n_u238out}
\end{figure*}

\begin{figure*}[htp]
\centering
\includegraphics[width = 0.90\textwidth]{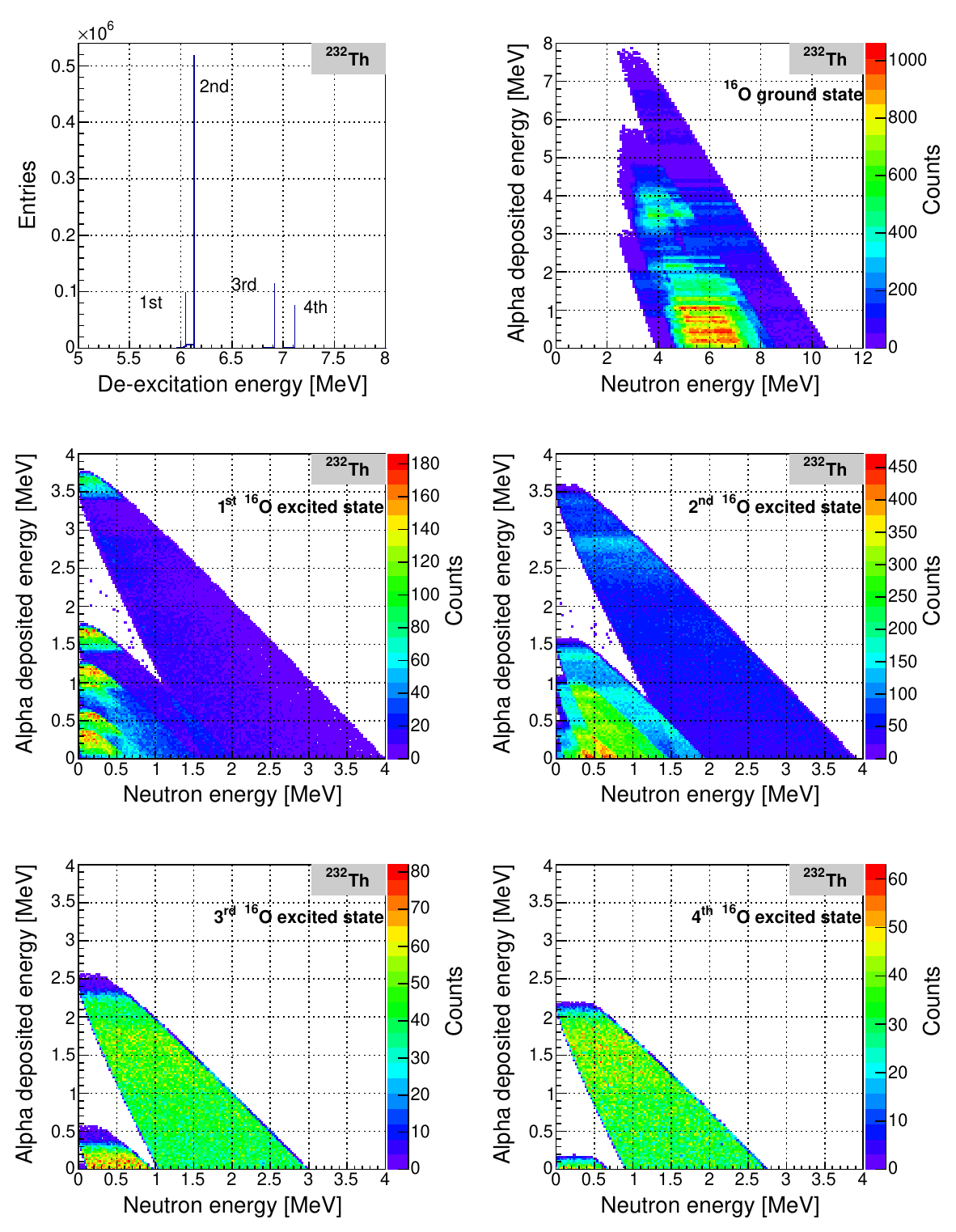}
\caption{SaG4n simulation of the $^{13}$C$(\alpha,\,n)^{16}$O reaction in the JUNO LS triggered by $\alpha$s from the $^{232}$Th chain in secular equilibrium. Top left: spectrum of $e^+e^-$ pairs and $\gamma$s from de-excitation of the 1$^{\mathrm{st}}$ and the 2$^{\mathrm{nd}}$ to  4$^{\mathrm{th}}$ \textsuperscript{16}O excited states, respectively. Other plots show correlations between the energy of emitted neutron and the deposited energy of $\alpha$ particle prior to its capture: top right plot for the case  when \textsuperscript{16}O is created in its ground state, while the remaining plots for the cases of the 1$^{\mathrm{st}}$ to  4$^{\mathrm{th}}$ \textsuperscript{16}O excited states, respectively. }
\label{fig:sag4n_th232out}
\end{figure*}

Figures~\ref{fig:sag4n_u238out} and \ref{fig:sag4n_th232out} show final states from the simulation of $^{13}$C$(\alpha,\,n)^{16}$O reactions triggered by $\alpha$s from the  $^{238}$U and $^{232}$Th chains in secular equilibrium, respectively. The complexity of the results is due to the extended number of emitted $\alpha$s of different energies in each of the decay chains. 

Numerical results regarding branching ratios of the energy levels of $^{16}$O and neutron yields $Y\lbrack n/\alpha \rbrack$ and $Y\lbrack n/\text{chain} \rbrack$ for $^{210}$Po, and the $^{238}$U and $^{232}$Th chains are summarized in table~\ref{tab:nY_v1_3}.

\begin{table*}[t]
\centering
\caption{Neutron yields and branching ratios (BR) of the populated \textsuperscript{16}O nucleus states from SaG4n simulations.}
\label{tab:nY_v1_3} 
\vskip 2pt
\begin{tabular}
{c | @{\hskip 10pt} c @{\hskip 10pt} c @{\hskip 10pt} c @{\hskip 10pt} c @{\hskip 10pt} c @{\hskip 10pt} | c @{\hskip 10pt} c @{\hskip 10pt} c} 
\hline
\hline
 chain or      &       &       & $BR$ [\%] &       &        & $Y\,\lbrack n/\alpha \rbrack$ & $\alpha$/chain & $Y\,\lbrack n/\text{chain} \rbrack$ \\
 $\alpha$ source  & $n_0$ & $n_1$ & $n_2$  & $n_3$ &  $n_4$ &               &                &         \\
\hline
$^{210}$Po      & 89.3  & 9.3   &  1.4   & 0.0   &  0.0   &  \num{5.11E-8}      &   1            &  \num{5.11E-8}  \\
$^{238}$U       & 51.5  & 7.9   & 29.3   & 7.0   &  4.3   &  \num{7.95E-8}      &   8            &  \num{6.36E-7}   \\
$^{232}$Th      & 43.9  & 8.5   & 34.2   & 8.1   &  5.3   &  \num{1.43E-7}      &   6            &  \num{8.58E-7}    \\
\hline
\hline
\end{tabular}
\end{table*}   

\subsection{Systematic effects}
\label{subsec:system}

Major systematic effects influencing the precision of our simulation of the $^{13}$C$(\alpha,\,n)^{16}$O reaction are the cross section uncertainties, comparison of our results on  neutron yield to SaG4n reference, and the choice of non-physical parameters assumed in the simulations. They are discussed in this section.

\subsubsection* {Alpha capture cross section}

The developers of the SaG4n software provided a comparison of the neutron yield from SaG4n to several calculation tools and nuclear data libraries for ($\alpha,\,n$) reactions. Their tests covered more than 10 types of target materials including pure carbon, using $^{235}$U, $^{238}$U, and $^{232}$Th decay chains as the $\alpha$ sources. The conclusion was that the neutron yields obtained with the SaG4n code and the JENDL/AN-2005 data library agreed with the experimental data within about 1\% for carbon and better than 10\% in most other cases~\cite{Men20, SaG4n_website}.

In 2018, Mohr re-evaluated the $^{13}$C$(\alpha,\,n)^{16}$O cross sections in the $\alpha$ energy region from 5 to \SI{8}{MeV}~\cite{Mohr18}, based on the capture data taken up to \SI{8}{MeV} from Harissopulos {\it et al.}~\cite{Har05}. Mohr proposed an average uncertainty of about 15\% in the cross section up to \SI{8}{MeV} in $\alpha$ energy.

\begin{figure*}[t]
\centering
\includegraphics[width = \textwidth]{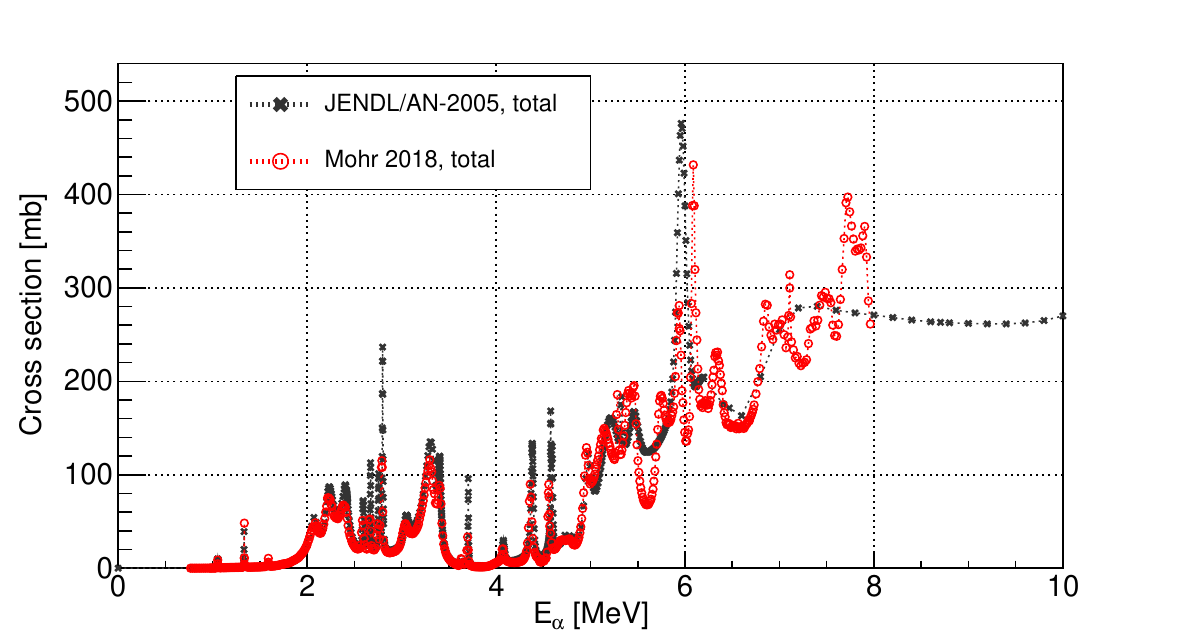}
\caption{Cross section of the $^{13}$C($\alpha$,\,$n$)$^{16}$O reaction as a function of $\alpha$s initial energy. Black x-markers show the cross section used in this work, as implemented in SaG4n from the JENDL/AN-2005 data library and shown also in figure~\protect\ref{fig:xs_and_a_energies}. Red markers (circles) represent the re-evaluation by Mohr in 2018~\protect\cite{Mohr18}, based on a 2005 experimental data set from Harissopulos et al.~\protect\cite{Har05}.}
\label{fig:alphan_xsec}
\end{figure*}

Figure \ref{fig:alphan_xsec} shows the $^{13}$C$(\alpha,\,n)^{16}$O cross sections as evaluated by Mohr and those available in the JENDL/AN-2005 library used in this work. The observed small discrepancy is mainly due to the use of different experimental data. However, thanks to the relatively close agreement of the curves in figure \ref{fig:alphan_xsec}, the total cross-sectional uncertainty of 15\% determined by Mohr was assumed to be the uncertainty in the neutron yield calculations using SaG4n. 

\subsubsection*{Comparison to SaG4n reference neutron yields}

The developers of SaG4n provided reference values of neutron yields from $(\alpha,\,n)$ reactions, using $^{235}$U, $^{238}$U, and \linebreak $^{232}$Th decay chains as the $\alpha$ sources, based on experimental data and their calculations~\cite{Men20, SaG4n_website}. The reference point on pure carbon target can be used to evaluate the precision of the neutron yields from this work.
We repeated our calculations for targets  with different hydrogen mass fractions in the range from zero (pure carbon target) to 50\%, while keeping the density of \SI{0.853}{g/cm^3}. Stated previously, the target representing the JUNO LS assumed a hydrogen mass fraction of 12.49\%.

\begin{figure*}[t]
\centering
\includegraphics[width = \textwidth]{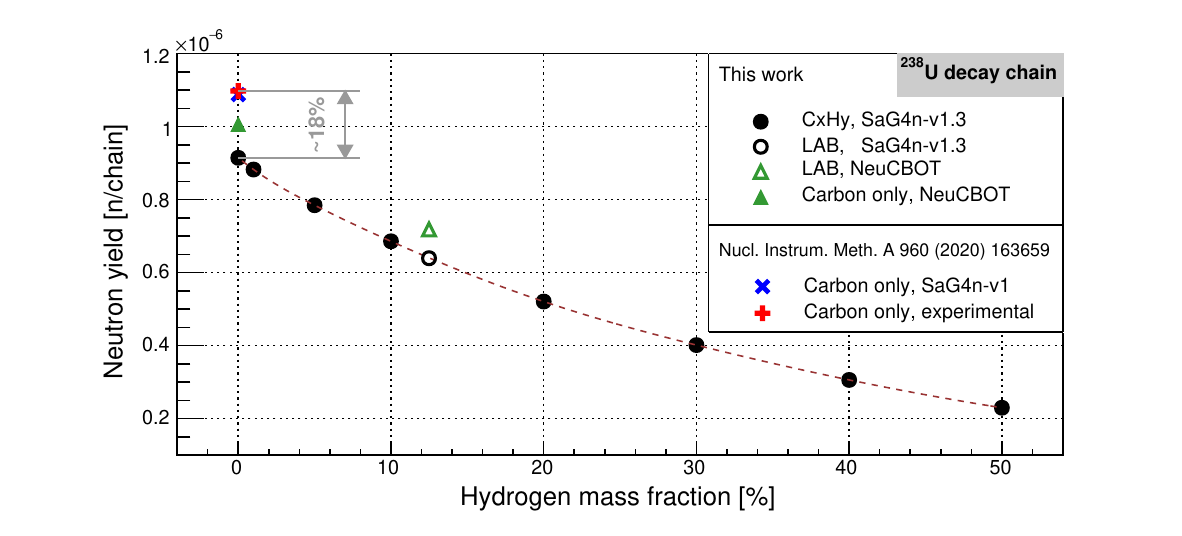}
\caption{Neutron yield as a function of hydrogen mass fraction in the target material for the $^{13}$C($\alpha$,\,$n$)$^{16}$O reaction, using the $^{238}$U decay chain as the $\alpha$ source. The black circles represent the results obtained with the tools utilized in this study: SaG4n v1.3 plus JENDL/AN-2005 with Geant4.11.1.2. The black open circle corresponds
to the respective yield for JUNO LAB. For comparison, calculations using the NeuCBOT framework with the JENDL/AN-2005 library are shown for the JUNO LAB case (green open triangle) and for pure carbon (green full triangle). The reference points taken from SaG4n developers~\protect\cite{Men20, SaG4n_website} for SaG4n v1.0 plus JENDL/AN-2005 with Geant4.10.4.p01 and for experimental data are shown in blue and red crosses, respectively. A deviation of approximately 18\% can be seen between the SaG4n result from this work and the only available experimental reference value for pure carbon.} 
\label{fig:alphan_CH_scan}
\end{figure*}

Figure~\ref{fig:alphan_CH_scan} compares our results for the \textsuperscript{238}U chain with different references points. The neutron yields obtained with SaG4n software version 1.3, using the JENDL/AN-2005 \linebreak data library, are shown for the LAB evaluation (black open circle) and for the variable hydrogen mass fractions (black full circles). We also show the pure-carbon reference values provided by SaG4n developers based on calculations with \linebreak SaG4n v1.0 and JENDL/AN-2005 (blue cross) and from measurements (red cross). While the two reference points are in agreement, a discrepancy of $\sim$18$\%$ can be seen with respect to our pure-carbon evaluation. The corresponding level of agreement for \textsuperscript{232}Th source was found to be 13$\%$.

For further comparison, simulations were also run using the NeuCBOT calculation framework~\cite{westerdale_radiogenic_2017, Gromov_2023, an_white_paper}. It can utilize the identical JENDL/AN-2005 cross-section database used in the SaG4n simulations. Two green triangle markers represent the \textsuperscript{238}U NeuCBOT results for a pure-carbon material as well as for LAB. Our calculations with NeuCBOT and SaG4n are consistent at the level of 10\% for \textsuperscript{238}U and 5\% for \textsuperscript{232}Th. 

The leading reason for these discrepancies was found to be due to the Geant4 version used during simulations. This work implemented the latest SaG4n software version 1.3, which was compiled with Geant4.11.1.2. The two reference values from the SaG4n article~\cite{Men20, SaG4n_website} were based on SaG4n software version 1.0, which was compiled with a modified version of Geant4.10.4.p01, and experimental data taken from~\cite{Fernandes_2017}, respectively. When we performed calculations with the older SaG4n version 1.1 and Geant4.10.05.p01, recommended by the authors, the yield difference compared to the original reference calculated yield (red marker), reduced to 8\%. This work assumes the latest software versions available at the time of writing. To account for the differences in yields between our latest results and the available experimental reference data, a systematic error of 18\% was assigned.

\subsubsection* {Maximum allowed step length for $\alpha$ simulation}

\begin{figure*}[t]
\centering
\includegraphics[width = \textwidth]{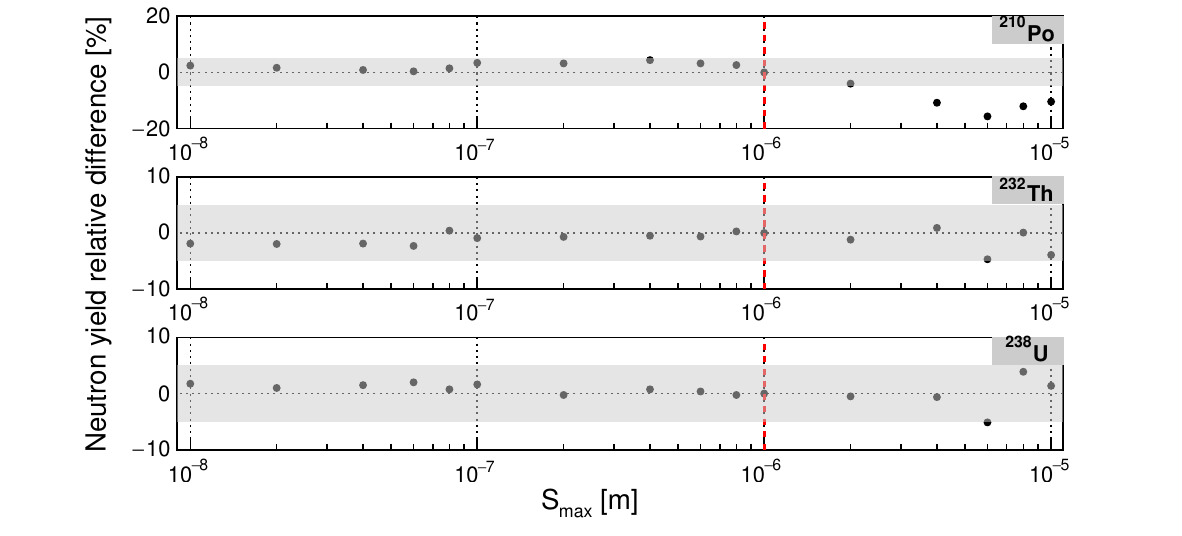}
\caption{Dependence of the neutron yield from $^{13}$C$(\alpha,\,n)^{16}$O reaction 
 on the maximum allowed step length $S_{\mathrm{max}}$ applied in the simulation of $\alpha$ particles with SaG4n. Y-axis shows relative differences with respect to the value of \SI{E-6}{m} used in this work and marked by the vertical red dashed line. The three different graphs show the results for $^{210}$Po (top) and for the $^{232}$Th (middle) and  $^{238}$U (bottom) chains. The three horizontal shaded areas represent the assigned $\pm$ 5\% systematic uncertainty due to $S_{\mathrm{max}}$.}
\label{fig:alphan_masl_scan}
\end{figure*}

The maximum allowed step length $S_{\mathrm{max}}$ of a simulated $\alpha$ in SaG4n (section \ref{subsec:sag4ninput}) was derived from the G4UserLimits class of the Geant4 standard library. Smaller values of $S_{\mathrm{max}}$ lead to more detailed tracking of the propagation of $\alpha$ particles, allowing for more precision on the yields, at the expense of longer computation times.
A study of the optimal $S_{\mathrm{max}}$ in the JUNO LS target was carried out by scanning the range of $S_{\mathrm{max}}$ from \SI{E-8}{m} to \SI{E-5}{m}, assuming $^{210}$Po as well as $^{238}$U and $^{232}$Th chains as the $\alpha$ sources. Figure~\ref{fig:alphan_masl_scan} shows the dependence of the neutron yield on the $\alpha$ step length. It can be seen that below $S_{\mathrm{max}} = \SI{E-6}{m}$, the yield approaches a stable value, within statistical fluctuations. Based on these studies, a value of $S_{\mathrm{max}} = \SI{E-6}{m}$ was assigned for the simulation results shown in this work. Regarding the impact of the $\alpha$ step length on the systematic uncertainty on the neutron yield, a 5\% value was assigned to reflect the fluctuations in the neutron yield seen for $S_{\mathrm{max}}$ smaller than \SI{E-6}{m}, for all three $\alpha$ sources.

\vspace{10mm}
   
\section{JUNO detector response to \texorpdfstring{$^{13}$C$(\alpha,\,n)^{16}$O}{13C(alpha, n)16O}}
\label{sec:genJunosw}

In this section, we discuss simulation of the JUNO detector's response to the products of $^{13}$C$(\alpha,\,n)^{16}$O reactions, described in the previous section. The JUNO collaboration has developed a dedicated Geant4-based software for detector performance studies, named JUNOSW~\cite{JUNOSW}. This package reflects a detailed detector geometry and models particle energy depositions, light production including non-linear quenching effects, light propagation, as well as the response of PMTs~\cite{Cao_Guofu_2022, PMTtest} and readout electronics. JUNOSW also includes event energy and vertex reconstruction algorithms. 
Further details regarding simulation and the reconstruction algorithms can be found in ~\cite{JUNOEnergyResPaper2024}. 

Section~\ref{subsec:sag4n_junosw} describes the interface between SaG4n and JUNOSW. The following section~\ref{subsec:ibd_cuts} treats coincident event selection identically to that used in the IBD event search in previous JUNO reactor antineutrino sensitivity studies~\cite{JUNONMO, JUNOSUBPERC}. In section~\ref{subsec:spectral_shapes} we finally present the spectral shapes of the IBD-like background due to the $^{13}$C$(\alpha,\,n)^{16}$O reaction expected in JUNO.

\subsection{SaG4n-JUNOSW interface}
\label{subsec:sag4n_junosw}

To simulate the detector response to $^{13}$C$(\alpha,\,n)^{16}$O events inside the JUNO LS, SaG4n outputs (section~\ref{subsec:sag4n_result}) were used to determine the initial particles with respective energies to be simulated with the JUNOSW. These particles include any products of the $^{16}$O de-excitation, either $\gamma$s or the $e^{+}e^{-}$ pair, and the emitted neutron. For simulation of $\alpha$ particles, which deposit only part of their initial energy in the LS before the capture ($E_{\mathrm{dep}}$), we apply an approximation that takes into account the energy dependence of the quenching effect in LS. The amount of emitted scintillation light for the same $E_{\mathrm{dep}}$ depends on the kinetic energy of $\alpha$ particle. Thus, we simulate the $\alpha$ with kinetic energy $E_{\mathrm{gen}}$, depositing all of its energy in LS chosen such, that the same amount of light would be produced as if the source $\alpha$ of higher energy would deposit $E_{\mathrm{dep}}$. As the amount of scintillation light emitted by the $\alpha$ prior to its capture is relatively small, this approximation was deemed appropriate.

All initial particles in the JUNOSW are isotropically generated from a single vertex. There is only one exception, which is the de-excitation from the first excited state of $^{16}$O, where the back-to-back topology of the $e^{+}e^{-}$ pair is considered, but at the same vertex again. The assumption of isotropy is acceptable for three reasons. Firstly, the $\alpha$ particle from the $\alpha$ decay is emitted isotropically. Secondly, although the outgoing neutron has some angular distribution with respect to the $\alpha$ particle direction at the moment of the reaction, this can be neglected for the following reasons. \linebreak $\alpha$ particles with kinetic energies below \SI{10}{MeV} only propagate in the LS up to $\sim\SI{100}{\micro\meter}$ within \SI{10}{\pico\second}, and the scintillation photons from the energy deposit are isotropic. Given the spatial and time resolutions of the detector, which are of the order of \SI{10}{cm} at the considered $\alpha$ particle energies and \SI{1}{\nano\second}, respectively, the JUNO detector sees the photons from the $\alpha$ energy deposition as from an isotropic point-like source. Therefore, the direction and track length of the $\alpha$ particle have no significant impact. Thirdly, since the ($\alpha,\,n$) reaction and the de-excitation of the daughter nucleus are independent processes, no angular correlation exists between the emitted neutron and the de-excitation photon (or $e^{+}e^{-}$ pair).

\subsection{IBD coincident event selection}
\label{subsec:ibd_cuts}

For the next step in the evaluation of backgrounds from the $^{13}$C$(\alpha,\,n)^{16}$O reaction expected in JUNO, we analyse  simulation results from JUNOSW after event reconstruction. We perform coincident event selection, same to the one used in the IBD event search in the reactor antineutrino analyses of JUNO~\cite{JUNONMO, JUNOSUBPERC}, namely we apply the following cuts:

\begin{compactitem}
    \item prompt-delayed time difference: $dT < \SI{1}{ms}$;
    \item prompt-delayed vertex distance: $dL < \SI{1.5}{m}$;
    \item radial fiducial volume cut on the prompt vertex: \\ $R_{\mathrm p} < \SI{17.2}{m}$;
    \item prompt reconstructed energy: $E_{\mathrm p} \in$ (0.7, 12.0)~MeV;
    \item delayed reconstructed energy: $E_{\mathrm d} \in$ (1.9, 2.5)~MeV or $E_{\mathrm d} \in$ (4.4, 5.5)~MeV.
\end{compactitem}

The efficiency $\mathcal{E}_{(\alpha,\,n)}^{\mathrm {IBD}}$ of these cuts is 0.84 for the $^{238}$U and $^{232}$Th chains and 0.87 for $^{210}$Po. The unequal efficiencies reflect the different energies of the respective \textsuperscript{16}O$^{*}$ de-excitation products, also having different propagation ranges in the LS, as it will be shown in the next section.

It is worth noting that the used criteria may be tuned, or their set might even be partly changed in the further analyses, which will be based on the collected data.

\subsection{\texorpdfstring{$^{13}$C$(\alpha,\,n)^{16}$O}{13C(alpha, n)16O} reconstructed energy spectra}
\label{subsec:spectral_shapes}

\begin{figure*}[htbp]
\centering
\includegraphics[width = 0.95\textwidth]{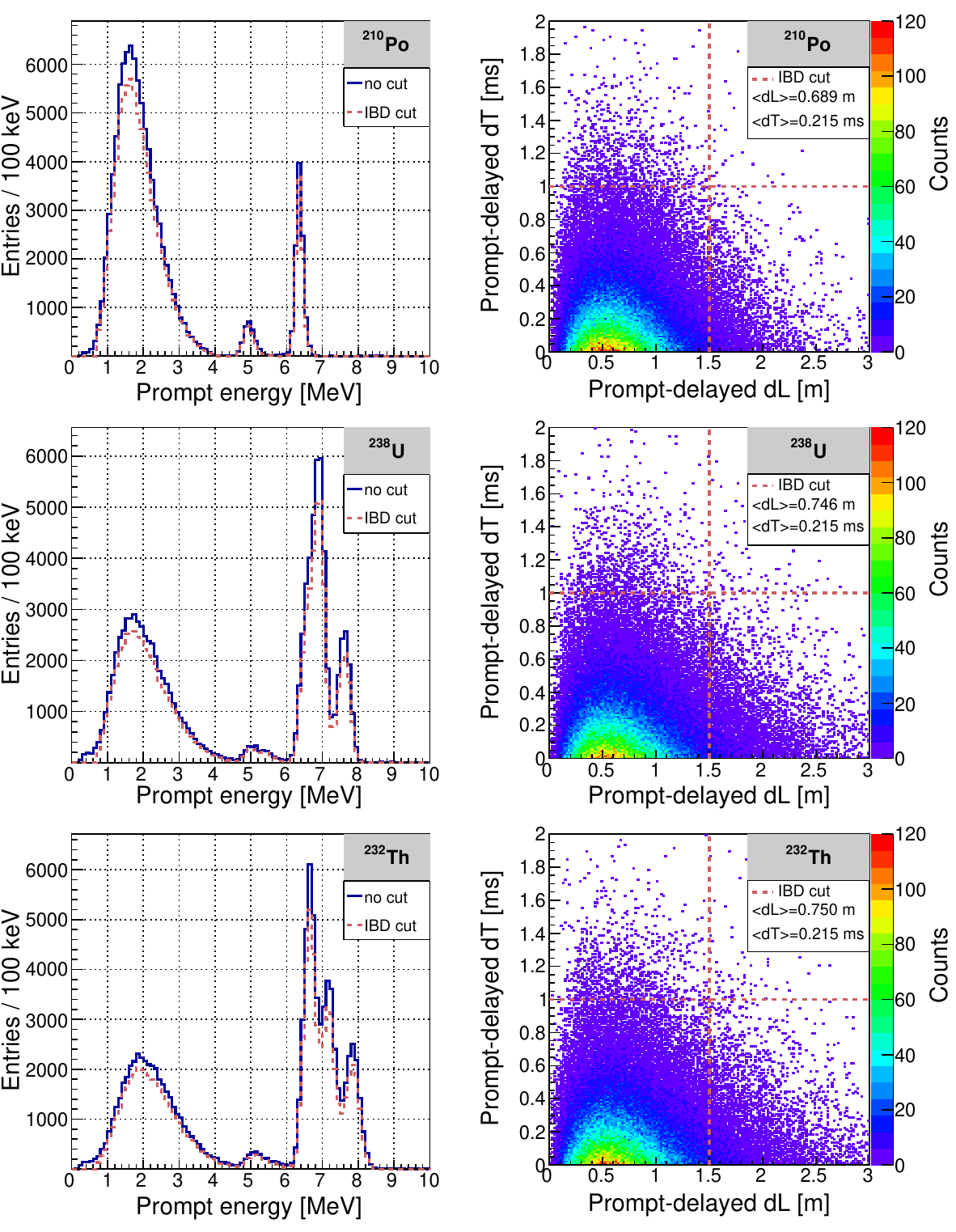}
\caption{Results of the $^{13}$C$(\alpha,\,n)^{16}$O simulation with JUNO software for different $\alpha$ sources: $^{210}$Po (top), $^{238}$U chain (middle), and $^{232}$Th chain (bottom). Left: The reconstructed prompt energy spectra before (solid blue line) and after (dashed red line) the IBD selection cuts. Right: The reconstructed prompt-delayed time $dT$ and distance $dL$. The red dashed lines demonstrate the applied IBD selection cuts.}
\label{fig:junosw_spectra}
\end{figure*}

The reconstructed energy spectra $E_{\mathrm p}$ representing the background in the antineutrino analysis are shown in the left column of figure~\ref{fig:junosw_spectra} for the $^{238}$U and $^{232}$Th chains and $^{210}$Po. Different structures seen in these spectra represent the three different mechanisms described in Sec.~\ref{sec:alphanProc} and in figure~\ref{fig:alphan_reac}. In all three spectra, the broad peak below $\sim$\SI{4}{MeV} reconstructed energy is due to protons scattered by neutrons \linebreak (Prompt-I). All other more narrow peaks are due to the de-excitation of nuclei. The peaks above $\sim$\SI{6}{MeV} are de-ex\-ci\-ta\-tion products of \textsuperscript{16}O$^{*}$ (Prompt-II), 
which have more complicated structure in case of $^{238}$U and $^{232}$Th, as $\alpha$s of higher energies from these chains, compared to the $\sim$\SI{5.3}{MeV} $^{210}$Po $\alpha$, can excite higher energy levels of \textsuperscript{16}O. The smallest peak seen around $\sim$\SI{5}{MeV} is due to the $\gamma$ from \textsuperscript{12}C$^{*}$ de-excitation (Prompt-III). We remind that additional energy depositions from proton recoil or $\alpha$ before its capture can modify the reconstructed prompt energy. This energy scale is also not corrected for the intrinsic non-linearity effects in LS and is anchored at a \SI{2.2}{MeV} $\gamma$ energy-scale equivalent. The right column of figure~\ref{fig:junosw_spectra} shows 2D distributions between the correlated reconstructed prompt-delayed time $dT$ and distance $dL$. The mean $dT$ of \SI{0.215}{ms} is the same for $^{238}$U and $^{232}$Th chains and $^{210}$Po. The mean $dL$ for $^{210}$Po of \SI{0.689}{m} is smaller than the mean $dL$ of \SI{0.746}{m} for $^{238}$U and \SI{0.750}{m} for $^{232}$Th due to different energies of \textsuperscript{16}O$^{*}$ de-excitation products with different ranges in LS.

\begin{table*}[t]
\centering
\normalsize
\caption{Rates of IBD-like background events due to $^{13}$C$(\alpha,\,n)^{16}$O reactions expected in JUNO from $^{238}$U and $^{232}$Th chains (minimal requirement for  the NMO measurement) and from the expected $^{210}$Po (from the $^{210}$Pb contamination and stand-alone). Label ``cpd'' stands for counts per day and CD refers to the whole JUNO LS volume. The considered fiducial volume is a sphere of \SI{17.2}{m} radius which corresponds to \SI{18.35}{kt} of LS.}
\label{tab:juno_rates} 
\vskip 2pt
\begin{tabular}{c | c c c c c c }
\hline
\hline 
sources & $Y_{n}$      & $c$   & $R_{\alpha}$ & $R_{(\alpha,\,n)}$ &  $\mathcal{E}_{(\alpha,\,n)}^{\mathrm {IBD}}$  &   $R_{(\alpha,\,n)}^{\mathrm {IBD}}$ \\
        & [$n$/chain] & [g/g] & [cpd/kt]             & [cpd/CD]   &                             & [cpd/FV] \\
\hline
$^{238}$U  & \num{6.36E-7}  &  \num{E-15}   &  1068  &  0.013  &   0.84  &  0.011 \\
$^{232}$Th & \num{8.58E-7}  &  \num{E-15}   &  352   &  0.006  &   0.84  &  0.005 \\
$^{210}$Pb/$^{210}$Po & \num{5.11E-8}  &  \num{5E-23}  &  12265 &  0.012  &   0.87  &  0.011 \\
\hline
$^{210}$Po & \num{5.11E-8}  & \num{3E-22}  &  70400 &  0.071  &   0.87  &  0.063 \\
\hline
\hline
\end{tabular}
\end{table*}   

\section{\texorpdfstring{$^{13}$C$(\alpha,\,n)^{16}$O}{13C(alpha, n)16O} event rates}
\label{sec:event_rates}

\subsection{Estimated \texorpdfstring{$^{13}$C$(\alpha, n)^{16}$O}{13C(alpha, n)16O} event rates from individual \texorpdfstring{$\alpha$}{alpha} sources} 

The $^{13}$C$(\alpha, n)^{16}$O event rates in the JUNO LS can be estimated in the following steps. For each individual source, we first evaluate the rate of $\alpha$ decays $R_{\alpha}$ in the LS,  assuming secular equilibrium in the decay chains:
\begin{equation}
R_{\alpha}~\left[\frac{\mathrm{cpd}}{\mathrm{kt}}\right] = c~\left[\frac{\mathrm{g}}{\mathrm{g}}\right]\cdot \frac{N_{\textrm{A}} \left[\displaystyle \frac{1}{\mathrm{mol}}\right]}{\tau~[\mathrm{day}] \cdot M~\left[\displaystyle \frac{\mathrm{g}}{\mathrm{mol}}\right]} \cdot 10^9~\left[\frac{\mathrm g}{\mathrm{kt}}\right]. 
\end{equation}

The rate $R_{\alpha}$ is expressed in cpd per \SI{1}{kt}. The expected concentration levels $c$ of $^{238}$U, $^{232}$Th, and $^{210}$Pb/$^{210}$Po are discussed in Sec.~\ref{sec:AlphaSource} and are expressed as the mass of mother isotope per gram of LS. The respective molar mass is $M$ and lifetime $\tau$, while $N_{\textrm{A}}$ is Avogadro's constant.  

\begin{figure*}[t]
\centering
\includegraphics[width = 0.7\textwidth]{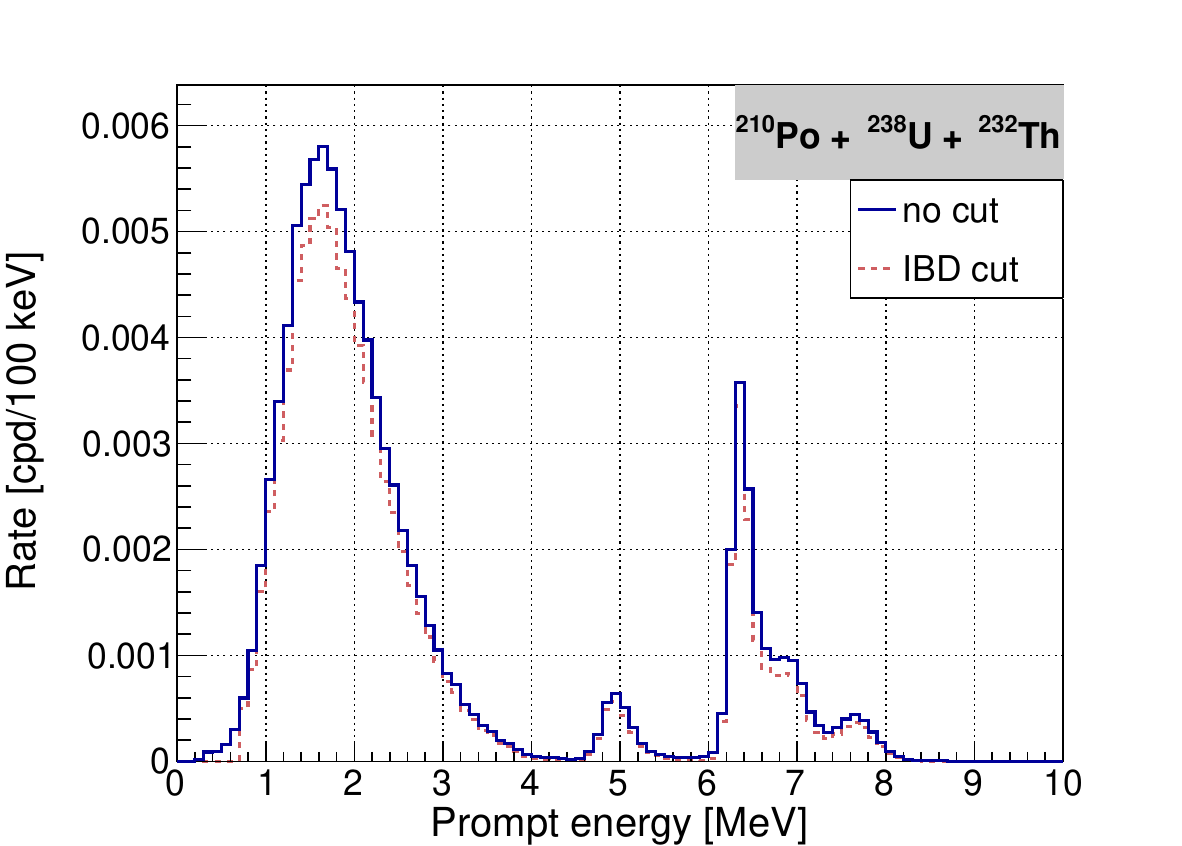}
\caption{The prompt reconstructed spectrum of the $^{13}$C($\alpha$,\,n)$^{16}$O reaction expected in JUNO before (solid blue line) and after (dashed red line) the IBD selection cuts, from the combined contributions of the $^{238}$U, $^{232}$Th, and $^{210}$Po $\alpha$ sources, with assumed rates summarised in table~\ref{tab:juno_rates}.} 
\label{fig:an_total}
\end{figure*}

In the second step, the expected rates $R_{(\alpha,\,n)}$ of \linebreak $^{13}$C$(\alpha,\,n)^{16}$O background events in the whole JUNO detector can be expressed as:
\begin{equation}
R_{(\alpha,\,n)} \left[\frac{\mathrm{cpd}}{\mathrm{CD}}\right] = R_{\alpha}~\left[\frac{\mathrm{cpd}}{\mathrm{kt}}\right] \cdot Y_{n}~\left[\frac{n}{\mathrm{chain}}\right] \cdot M_{\text{LS}}~[\mathrm{kt}], 
\end{equation}
where $Y_n$[$n$/chain] are the neutron yields per chain from table~\ref{tab:nY_v1_3} and $M_{\text{LS}}$ is the \SI{20}{kt} mass of the JUNO LS. Finally, taking into account the efficiencies $\mathcal{E}_{(\alpha,\,n)}^{\mathrm {IBD}}$, {\it i.e.}, the probability that the $(\alpha,\,n)$ reaction passes the IBD selection criteria, we express the final background rates $R_{(\alpha,\,n)}^{\mathrm {IBD}}$ in the antineutrino measurement in the spherical FV of \SI{17.2}{m} radius (\SI{18.35}{kt}) due to $^{13}$C$(\alpha, n)^{16}$O reactions in JUNO. Rates of \SI{0.011}{cpd/FV} and \SI{0.005}{cpd/FV} are expected from the $^{238}$U and $^{232}$Th chains in secular equilibrium, respectively. The dominant contribution of \SI{0.063}{cpd/FV} is evaluated \linebreak from unsupported $^{210}$Po and an additional \SI{0.011}{cpd/FV} from the $^{210}$Po from $^{210}$Pb that is out of equilibrium with the $^{238}$U chain. All the ingredients for this calculation are summarized in table~\ref{tab:juno_rates}. The overall
$^{13}$C$(\alpha,\,n)^{16}$O background expected in JUNO amounts to \SI{0.090}{cpd/FV} and its shape is shown in figure~\ref{fig:an_total}.

\subsection{Event rate uncertainties}
\label{subsec:rate_uncertainty}

In this section, we evaluate the sources of uncertainty due to detector response and characteristics. In Sec.~\ref{sec:bipo} we discuss the precision with which the realistic contamination of the LS with $\alpha$ emitters can be determined. In Sec.~\ref{sec:quneching} we evaluate the impact of the accuracy of the JUNO LS quenching  effect. Section~\ref{sec:total_sys} summarizes all effects to provide an estimation of the total systematic uncertainty on our results, taking into account also the uncertainties presented in section~\ref{subsec:sag4n_result} regarding the simulation of the $^{13}$C$(\alpha,\,n)^{16}$O reaction with SaG4n.

\subsubsection{Evaluating \texorpdfstring{$\alpha$}{alpha} source concentration}
\label{sec:bipo}

Table \ref{tab:juno_rates} summarises the expected measurable ($\alpha,\,n$) event rates according to the assumed $\alpha$ source concentration levels within the LS. Therefore, the uncertainty in the predicted ($\alpha,\,n$) rate depends directly on the uncertainty in the measured radioactivity concentration levels within the detector. A commonly used \textit{in-situ} method to extract the concentration of the \textsuperscript{238}U and \textsuperscript{232}Th chains in secular equilibrium~\cite{BorexinoBGs,SNO+Solar}, is through the rate measurement of their daughter decay pairs \textsuperscript{214}Bi-\textsuperscript{214}Po and \textsuperscript{212}Bi-\textsuperscript{212}Po, respectively. These Bi-Po event pairs consist of the $\beta$-decay of Bi, followed rapidly by the $\alpha$-decay of Po, providing a possibility of coincident event tagging with high efficiency and purity. These samples also provide excellent data for tuning the $\alpha/\beta$ discrimination methods~\cite{Basilico2023}, that are also being implemented in JUNO~\cite{Rebber2021}.

The amount of out-of-equilibrium \textsuperscript{210}Po can be identified directly via application of these $\alpha/\beta$ discrimination methods~\cite{BOREXINOGEO}. In this work, we assume that JUNO data will allow extracting the precision of the $\alpha$ emitters in the LS with an uncertainty of 5\%.

\subsubsection{Scintillation quenching factors}
\label{sec:quneching}

JUNOSW models the quenching effects in the LS energy response following the semi-empirical Birks' law \cite{Birks1951}, with three coefficients $kB$, defined for $e^{+}/e^{-}$, protons, and $\alpha$s. The values of $kB$ used in this work were assumed from measurements made by the Daya Bay experiment, where more details can be found in \cite{JUNOEnergyResPaper2024}. The thorough calibration of the quenching parameters in JUNO is planned based on deployable source calibration~\cite{JUNOCALIB} and ongoing table-top experiments. 

In this work, uncertainties in the proton quenching factors directly impact the low energy part of the prompt \linebreak $^{13}$C$(\alpha,\,n)^{16}$O spectrum. To determine the level at which the proton quenching uncertainty can impact the spectrum, we varied the Birks' coefficients within a range of $\pm$10\% in the simulation of $^{241}$Am-$^{13}$C neutron calibration source~\cite{Liu2015}. We performed multiple simulations of this source placed at the detector's center, accounting for its detailed geometry. For each simulation, the reconstructed prompt energy spectrum was produced, applying the same IBD analysis cuts defined in section~\ref{subsec:ibd_cuts}. It was determined that the peak position of the low energy proton recoil peak can be defined with a precision of $\sim$$1\%$.

The precision of the $\alpha$ quenching factor has limited impact on the ($\alpha,\,n$) background. In order to evaluate it, we repeated our simulations by varying the $kB$ values of $\alpha$s by $\pm$5\%, i.e. the precision certainly worse that JUNO expects to achieve on this parameter. The resulting changes in the ($\alpha,\,n$) prompt reconstructed energy spectrum were found to be less than 1\%. 

Overall for this work, a 5\% conservative uncertainty was assigned to the $^{13}$C$(\alpha,\,n)^{16}$O event rates due to the quenching factors of protons and $\alpha$s. 

\subsubsection{Summary of \texorpdfstring{$^{13}$C$(\alpha,\,n)^{16}$O}{13C(alpha, n)16O} event rate uncertainty}
\label{sec:total_sys}

The sources of systematic uncertainties for the $^{13}$C$(\alpha,\,n)^{16}$O event rates, following the above discussions, are sum\-ma\-ri\-sed in table~\ref{tab:rate_uncertainty}. The total value of 25\% is calculated as \linebreak the quad\-ra\-tic sum, conservatively neglecting possible correlations among different sources.  

\begin{table}[htbp]
\centering
\caption{Summary of the uncertainties of the estimated $^{13}$C$(\alpha,\,n)^{16}$O event rates.}
\label{tab:rate_uncertainty} 
\vskip 2pt
\begin{tabular}{c  c}
\hline
\hline
Uncertainty source &   Relative uncertainty \\
\hline
SaG4n reference value discrepancy &  18\%   \\
$^{13}$C$(\alpha,\,n)^{16}$O cross section           &  15\%    \\
$\alpha$ maximum step length dependence         &   5\%    \\
Detector response       &   5\%     \\
Radioactivity concentration     &   5\%     \\
\hline
Total (quadratic sum)              &   25\%     \\
\hline
\hline
\end{tabular}
\end{table}   
   
\section{Conclusions}
\label{sec:conclusion}

The $^{13}$C$(\alpha,\,n)^{16}$O reaction represents an important \linebreak back\-ground in the detection of electron antineutrinos in LS detectors, as for the cases of reactor and geoneutrinos. This work has presented the first specific evaluation of this interaction in JUNO, using novel techniques and implementing the expected radiopurity of its LS. In particular, we have applied the SaG4n simulation tool version 1.3 and the current version of the JUNO simulation and reconstruction software. The total expected rate is $0.090 \times (1 \pm 0.25)\,\mathrm{cpd}$ in the fiducial volume of the analysis (\SI{18.35}{kt}) from $^{232}$Th and $^{238}$U chains and additional out-of-equilibrium $^{210}$Po, as in table~\ref{tab:juno_rates}. The expected shape of this background is shown in figure~\ref{fig:an_total}. According to recent calculations~\cite{JUNONMO}, the estimated IBD rate is equal to \SI{47.1}{cpd}. This is much higher than the $^{13}$C$(\alpha,\,n)^{16}$O background. Thus, given the expected levels of $\alpha$ contamination in the JUNO LS, the impact on the reactor and geoneutrino measurements is found to be minimal, provided this background is appropriately constrained during the analysis.

While this evaluation has been performed specifically for JUNO using LAB-based LS, our results can be exploited also for other LS-based experiments. Particularly useful can be the provided neutron yields in table~\ref{tab:juno_rates} and supplementary material available online regarding SaG4n simulation configurations and results for $^{232}$Th, $^{238}$U, and $^{210}$Po, as well as the dependence of our results on the hydrogen mass fraction of the LS.  And last but not least, this work employs one particular nuclear database, JENDL/AN-2005. Additional calculations using other newer libraries may be needed in the future. It will mainly help to more precisely evaluate the associated systematic uncertainties. Moreover, evaluation of the actual background levels based on the JUNO data is also planned.

\section*{Acknowledgments}
\label{sec:acknowledgments}

This work was supported by the Chinese Academy of Sciences, the National Key R\&D Program of China, the Guangdong provincial government, the Tsung-Dao Lee Institute of Shanghai Jiao Tong University in China, the Institut National de Physique Nucléaire et de Physique de Particules (IN2P3) in France, the Istituto Nazionale di Fisica Nucleare (INFN) in Italy, the Fond de la Recherche Scientifique (F.R.S-FNRS) and the Institut Interuniversitaire des Sciences Nucléaires (IISN) in Belgium, the Conselho Nacional de Desenvolvimento Científico e Tecnológico in Brazil, \linebreak the Agencia Nacional de Investigacion y Desarrollo and \linebreak ANID - Millennium Science Initiative Program~-- \linebreak ICN2019\_044 in Chile, the European Structural and Investment Funds, the Czech Ministry of Education, Youth and Sports and the Charles University Research Centerin Czech Republic, the Deutsche Forschungsgemeinschaft (DFG), the Helmholtz Association, and the Cluster of Excellence \linebreak PRISMA+ in Germany, the Joint Institute of Nuclear Research (JINR) and Lomonosov Moscow State University in Russia, the Slovak Research and Development Agency in Slovak, MOST and MOE in Taipei, the Program Management Unit for Human Resources \& Institutional Development, Research and Innovation, Chulalongkorn University, and Suranaree University of Technology in Thailand, the Science and Technology Facilities Council (STFC) in the UK, University of California at Irvine and the National Science Foundation in the US, 
and the State project ``Science'' by the Ministry of Science and Higher Education of the Russian Federation under the contract 075-15-2024-541.

\bibliographystyle{spphys}
\bibliography{alpha_n_JUNO_2025-05-01}

\end{document}